\documentclass[aps,prb,reprint,amsmath,twocolumn,superscriptaddress,showpacs]{revtex4-1}

\usepackage{amssymb}
\usepackage{amsmath}
\usepackage{ulem}
\usepackage{verbatim}
\usepackage{color}
\usepackage{graphicx}
\usepackage{grffile}
 \usepackage[ampersand]{easylist}
\usepackage{bm}
\usepackage{bbm}
\usepackage[caption=false]{subfig}
\usepackage{cancel}
\usepackage{soul}

\def\ket#1{|#1\rangle}

\def\cexp#1{\langle#1\rangle}

\def\d{\mathrm{d}}

\def\pdev#1#2{\frac{\partial#1}{\partial#2}}
\def\dev#1#2{\frac{\mathrm{d}#1}{\mathrm{d}#2}}

\newcommand{\bl}{\begin{easylist}}
\newcommand{\el}{\end{easylist}}

\let\oldsqrt\sqrt
\def\sqrt{\mathpalette\DHLhksqrt}
\def\DHLhksqrt#1#2{%
\setbox0=\hbox{$#1\oldsqrt{#2\,}$}\dimen0=\ht0
\advance\dimen0-0.2\ht0
\setbox2=\hbox{\vrule height\ht0 depth -\dimen0}%
{\box0\lower0.4pt\box2}}

\begin{document}
\author{Philip J D Crowley}
\affiliation{London Centre for Nanotechnology, University College London,
Gordon St, London, WC1H 0AH, United Kingdom}
\author{A G Green}
\affiliation{London Centre for Nanotechnology, University College London,
Gordon St, London, WC1H 0AH, United Kingdom}
\title{An Anisotropic Landau-Lifschitz-Gilbert model of dissipation in qubits}

\begin{abstract}
We derive a microscopic model for dissipative dynamics in a
system of mutually interacting qubits coupled to a thermal bath that generalises the
dissipative model of Landau-Lifschitz-Gilbert to the case of 
anisotropic bath couplings. 
We show that the dissipation acts to bias the quantum trajectories towards a
reduced phase space.
This model applies to a system of superconducting flux qubits whose coupling to
the environment is necessarily anisotropic. We study the model in the context
of the D-Wave computing device and show that the form of environmental coupling
in this case produces dynamics that are closely related to several models
proposed on phenomenological grounds.
\end{abstract}

\maketitle

\section{Introduction}
The Landau Lifschitz-Gilbert-equation \cite{landau1935,gilbert2004} provides
a phenomenologically motivated description of the stochastic, dissipative
evolution of a spin system. Conceived as a model for an open magnetic system,
the dynamics consists of two terms corresponding to precessing, Hamiltonian
evolution, and noisy, relaxing, dissipative evolution. This form of dissipation
corresponds to the classical limit of a spin system coupled isotropically to a
bosonic bath. 

The behaviour of a Superconducting quantum device can be mapped to spin
dynamics~\cite{leggett1987}. In the case of a flux qubit, the low energy dynamics map on to those of a
spin-$1/2$ magnetic moment, and an isolated qubit will exhibit only the
Larmor precession described by a dissipation free Landau-Lifschitz-Gilbert
equation; the qubit wavefunction evolves such that the absolute amplitudes of
the two energy eigenstates remain constant whilst the phase between them changes
at a constant rate.

This relationship suggests that qubits coupled to the environment may
display the same dissipative behaviour as magnetic moments, governed by the full
dissipative Landau-Lifschitz-Gilbert equation. This model has been used to model
an extended array of superconducting qubits in Ref~\onlinecite{vinci2014} whilst
related vector models have been used in
Refs~\onlinecite{smolin2013,wang2013,shin2014}.

However, the Landau-Lifschitz-Gilbert equation 
describes the dissipative dynamics of a two-level system that
is exposed to isotropic environmental coupling, i.e. identical
baths coupled to the $\hat{s}_x$, $\hat{s}_y$, and
$\hat{s}_z$ operators. For qubits, these operators may have different
physical origins and hence will couple differently to
noise. Thus, the dissipation will also be anisotropic, as the stochastic noise
and dissipative terms are related by the fluctuation dissipation relation. As a
result, systems, such as flux qubits, with anisotropic couplings have
corresponding anisotropies in the dissipation and noise.

Due to the physical geometry of the superconducting flux qubit, stray flux and other environmental effects couple to the $\hat{s}_z$ operator. The anisotropy of this environmental coupling introduces qualitatively new features into the system's dynamics. We show the
existence of a regime where the qubit dynamics are typically confined to a low
dimensional sub-manifold of the full Hilbert space. 

The dissipative dynamics of a two level system has been studied
extensively~\cite{leggett1987}. In this work, we use a Keldysh path
integral~\cite{kamenev2011,orth2010,green2006} to derive a
Langevin~\cite{schmid1982,weiss1999} description of the dynamics of a flux qubit accounting for the anisotropic coupling to the environment. This extends
the previous work on dissipative spin
models~\cite{Kubo1970,hohenberg1977,jayannavar1991brownian,garanin1990dynamics,Garanin1991,plefka1993,garanin1997fokker,nussinov2005spin,katsura2006voltage,chudnovsky2012conservation}.

An accurate model for the
dissipative dynamics of a flux qubit can be used to assess the capabilities 
of putative quantum technologies. We apply our model to the D-Wave computing
machine, which consists of a large array of controllable flux qubits. Extensive
analysis has sought to correlate the behaviour of this machine with various
quantum and classical
models~\cite{boixo2014,smolin2013,wang2013,shin2014,vinci2014,shinagain2014,crowley2014,bauer2015}.
Since classical dynamics correspond to a particular restriction upon
fully quantum dynamics, the effectiveness of this approach is dependent upon
identifying the appropriate restrictions that
correspond to the classical limit. We show in an appropriate strong
coupling limit the biasing of trajectories in the anisotropic Langevin
equation allows one to obtain dynamics reminiscent of the heuristic models of
Refs~\onlinecite{smolin2013,shin2014,shinagain2014}.

\section{The Anisotropic dissipation of flux qubits}
A superconducting flux qubit will couple to its environment in various
ways. The environmental degrees of freedom may consist of charge fluctuations,
flux noise, coupling to nearby spins, or trapped vortices. When
it is not possible to interrogate such environmental subsystems, the
information contained in the state of these subsystems is lost from the system
of interest\textemdash the qubit. This decoherence inhibits the ability of a
qubit to remain in a given state indefinitely with good fidelity, and in turn
limits the ability of an array of qubits to sustain entanglement.

This loss of unitary evolution generates dissipative dynamics that,
over sufficiently long times, drive the system towards certain equilibrium states,
or dynamical fixed points. In this sense, the effects of decoherence are
inherently inhomogeneous over the system's Hilbert space, they affect different
states in different ways, driving them towards different fixed points or along
different paths to such fixed points.

We anticipate that a system exchanging energy with an environmental bath will
relax to a fixed point given by the Gibbs state. There remain, however, a plurality of dynamics
that result in the system relaxing to this state, and we expect that
different baths and couplings with different physical origins will result in
different relaxation dynamics. This is relevant to controlled quantum systems
in which the dynamics of interest occur before the system has relaxed to
thermal equilibrium, but not necessarily on timescales where dissipation can
be neglected.

The Landau-Lifschitz-Gilbert describes the dissipative
dynamics of a two-level system with isotropic coupling to the environment. It
is often quoted in one of two equivalent forms;
\begin{subequations}
\begin{align}
\mathbf{\dot{s}} + \mathbf{s} \times \left[\left(\mathbf{B} +
\bm{\eta}\right) - \gamma \mathbf{\dot{s}}\right] = 0, 
\label{eq:LLG} \\ 
\mathbf{\dot{s}} + \frac{1}{1+ s^2
\gamma^2}\mathbf{s} \times \left[
\left(\mathbf{B} + \bm{\eta}\right) + \gamma \,\mathbf{s} \times \left( 
\mathbf{B} + \bm{\eta} \right)\right] = 0.
\label{eq:LLG2}
\end{align}
\end{subequations}
These are non-linear differential equations in a vector $\mathbf{s}$ of
magnitude $s$ that parametrises a qubit spin coherent state $\mathbf{\hat{s}}
\ket{\mathbf{s}} = \mathbf{s} \ket{\mathbf{s}}$.  The equation is stated here in
terms of an effective magnetic field, $\mathbf{B}$. More generally, the
magnetic field may be replaced by an appropriate derivative of the spin
Hamiltonian, ${\bf B}=-\nabla_{\mathbf{s}}H(\mathbf{s})$. $\bm{\eta}$ describes
a stochastic noise that satisfies the fluctuation dissipation relation;
$\cexp{\eta_\alpha(\omega) \eta_\beta(\omega')} = \gamma \omega
\coth\left(\frac{\omega}{2 T}\right) \delta_{\alpha \beta}
\delta(\omega+\omega')$~\footnote{Neglecting zero-point fluctuations of the
environment this reduces to the time domain form
\unexpanded{$\cexp{\eta_\alpha(t) \eta_\beta(t')} = 2 \gamma T \delta_{\alpha
\beta} \delta(t-t')$}.}.

This description closely resembles the phenomenological Bloch equations used
to model nuclear magnetization, the difference being the
Landau-Lifschitz-Gilbert equation describes a single spin and not an
ensemble average, thus the decay must preserve the spin $|\mathbf{s}|=s$.
Neglecting zero-point fluctuations, this Langevin equation describes the
dissipative evolution of a single qubit in the presence of a magnetic field,
$\mathbf{B}$, and coupled isotropically to its environment.

The dynamics of a system of multiple interacting qubits may be described by a set
of coupled Landau-Lifschitz-Gilbert equations. These dynamics are general
and constitute a restriction to product states. Coherent dynamics are
permitted for individual spins in these states, but there is no entanglement
between spins. In this sense the equations correspond to the classical limit of the
system.

Although originally developed to describe spins that are isotropically
susceptible to noise, the Landau-Lifschitz-Gilbert equations have been used to
model the classical dynamics of dissipative qubits~\cite{vinci2014}, whilst
other authors have also made use of similar vector
models~\cite{shin2014,shinagain2014}. However, as argued above, the physics
underlying superconducting flux qubits implies that noise and dissipation will
be anisotropic. Whilst the microscopic origins of flux qubit decoherence are not
fully understood~\cite{mcdermott2009}, many environmental interactions are often
modelled using linear $\hat{s}_z$
couplings.\cite{leggett1987} These include flux noise and coupling to impurity
spin~\cite{prokof2000,faoro2006,desousa2007,mcdermott2009} or
boson~\cite{shnirman2002,dube2001} degrees of freedom, and have been
shown to be significant contributions to decoherence~\cite{yoshihara2006}.
Whilst in the context of the D-Wave computing machine it has been argued that
linear $\hat{s}_z$ coupling is the correct minimal model \cite{Amin2009}.
Without loss of generality any linear coupling can be chosen to be $\hat{s}_z$
and whether longitudinal or transverse.
Due to the close relationship between the damping and noise terms enforced by
the fluctuation dissipation relation, this cannot be remedied solely by an
appropriate adjustment to the noise term; an adjustment to the noise term
effects the damping to term in a manner that is difficult to guess.

\section{A Dynamical Equation with Anisotropic Dissipation}
\label{sec:Derivation}
Here we follow the approach of Ref~\onlinecite{kamenev2011} to find the
partition function $Z = \prod_i \int D[\mathbf{s}_i] \mathrm{e}^{i
S[\{\mathbf{s}_i]\}}$ for an open quantum system, and evaluate the dynamics of a
qubit using a stationary phase approximation.
In a closed system, the action of the spins is given by
\begin{equation}
S_0 = \int_\mathcal{C} \mathrm{d}t\, L_0 = \int_\mathcal{C} \mathrm{d}t 
\left(\sum_i \dot{\mathbf{s}_i}\cdot \mathbf{A}_i - H(\{\mathbf{s}_i\}) \right),
\label{eq:bareaction}
\end{equation}
where $\mathbf{A}_i = \frac{1-\cos
\theta_i}{\sin \theta_i} \hat{\bm{\phi}}$ is the single monopole vector
potential.
This action describes the precession of a spin around the axis of the
magnetic field. When the spin is coupled to a bath of oscillators through the
$s_z$-component, the action becomes
\begin{equation}
\label{eq:fullaction}
S = S_0 + \sum_{i,\alpha} \int_\mathcal{C} \mathrm{d}t \left[s_{i,z}
g_{i,\alpha} x_\alpha + \frac{m}{2}\left(\dot{x}_\alpha^2 - \omega_\alpha^2
x_\alpha^2\right)\right].
\end{equation}
We use Keldysh field theory to find the dynamics of this system. This 
provides a methodology for treating an open quantum system, enforcing the
necessary fluctuation dissipation relation between the stochastic noise and	
deterministic damping terms induced by decoherence.

A state vector is sufficient to define the behaviour of a closed system. 
Non-equilibrium, open quantum systems require an ensemble of state
vectors, or a density matrix, because of the loss of information to the bath.
Evolving a density matrix $\hat{\rho}$ requires both pre- and 
 post-multiplication by the time evolution operator $\hat{U}(t_1,t_2) =
\mathcal{T} \exp{\left[i \int_{t_2}^{t_1} \mathrm{d} t \, \hat{H}(t)\right]}$. 
Thus $\hat{\rho}_t = \hat{U}(t,0) \hat{\rho}_0 \hat{U}(0,t)$ in contrast to the
time evolution of a state vector which requires
only one time evolution operator $\ket{\psi_t} = \hat{U}(t,0) \ket{\psi_0}$.
This gives rise to a doubling of the degrees of freedom in the Keldysh theory
of open systems compared to those required to describe the evolution of a
closed system. These degrees of freedom correspond to the forward and backward
branches of the Keldysh contour, $\mathcal{C}$, over which
Eqs~\eqref{eq:bareaction} and \eqref{eq:fullaction} are integrated.

It is useful in calculations to separate diagonal and off-diagonal
contributions to the density matrix. The former are described by the sum of
fields on the forwards and backwards Keldysh contour, and the latter by their
difference.
These fields are usually called the classical and quantum fields respectively.
Expanding in the quantum fields generates a series of higher order quantum
corrections to the classical dynamics. Performing this rotation, and expanding to first order in
the quantum fields results in each spin having a total action $S = S_0 +
S_\mathrm{diss}$
\begin{widetext}
\begin{equation}
\label{eq:rotatedaction}
S = 
\int_{\mathbb{R}} \mathrm{d}{t} 
\left[ \sum_i
2\, \mathbf{s}_i^{q} \cdot \left( \pdev{L_0}{\mathbf{s}_i} - \dev{}{t}
\pdev{L_0}{\dot{\mathbf{s}_i}} \right) + 2 \sum_{i,\alpha} g_{i,\alpha} \left(
s_{i,z} x_\alpha^q + s_{i,z}^q x_\alpha \right) + \sum_\alpha
\begin{pmatrix} x_\alpha & x_\alpha^q \end{pmatrix}  
\begin{pmatrix} {0} & {[D_\alpha^A]^{-1}} \\ {[D_\alpha^R]^{-1}} &
{[D_\alpha^{-1}]^{K}} \end{pmatrix}
\begin{pmatrix} x_\alpha \\ x_\alpha^q \end{pmatrix}
\right].
\end{equation}
\end{widetext}
The first term in Eq~(\ref{eq:rotatedaction}) encodes the Hamiltonian
(closed system) dynamics of each spin. The remaining terms encode the
dissipative dynamics induced by interactions with the bath. 

\subsection{A Langevin Equation}
\label{sec:langevin}
The dynamical equation can be obtained from a limit of the
Keldysh Theory as follows: In an open system the state of bath cannot be
interrogated, thus it is integrated out to obtain the dissipative contribution
to the action in terms of the spin alone;
\begin{equation}
\label{eq:sdiss}
S_{\mathrm{diss}} = - \sum_{i} \int_{\mathbb{R}} \mathrm{d}{t}
\left[ 
\left(
\begin{array}{cc}
s_{i,z} & s_{i,z}^q \\
\end{array}
\right) 
\left(
\begin{array}{cc}
0 & {D^A} \\
{D^R} & D^{K} \\
\end{array}
\right)
\left(
\begin{array}{c}
s_{i,z} \\
s_{i,z}^q
\end{array}
\right) 
\right]
\end{equation}
where the correlators ${D^A} = \sum_\alpha g_{i,\alpha}^2 D_\alpha^A$, and
similarly for $D^R$ and $D^K$, are equal for each spin, and defined by the bath
spectral function $J(\omega) = \sum_\alpha
\pi g_{i,\alpha}^2/\omega \,\delta\left(\omega-\omega_\alpha\right)$.
The equation is made linear in the spin fluctuation variables $s_z^q$ by a
Hubbard-Stratonovich transformation
\begin{equation}
S_\mathrm{diss} = \sum_i \int_{\mathbb{R}} \mathrm{d}{t} 
 \left( 2 \,s_{i,z}^q D^R s_{i,z} + 2 s_{i,z}^q \eta_i - \eta_i
 \left[D^K\right]^{-1}\eta_i \right)
\end{equation}
which introduces the variable $\eta$ that later appears as the stochastic term
in the dynamical equations. $\mathbf{s}^q$ now act as Lagrange multipliers
and can be integrated out of the partition function to obtain
\begin{equation}
\begin{split}
\label{eq:PartFunc}
Z &= \prod_i \int D[\eta_i] \mathrm{e}^{- i \int \eta_i [D^K]^{-1} \eta_i \,
\mathrm{d}{t}}\int D [\mathbf{s}_i] \, \times \\ & \delta\left(
\frac{1}{s^{2}}
\mathbf{s}_i
\times
\dot{\mathbf{s}}_i + \nabla_{\mathbf{s}_i} H  + \int_{\mathbb{R}} \d t'
D^R(t-t') s_{i,z}(t') \mathbf{z} + \eta_i \mathbf{z} \right)
\end{split}
\end{equation} 
the dynamical equations describing the semi-classical dynamics of the system are
then found are then found by taking the saddle points of the path integral. The
resulting dynamical equation is 
\begin{equation}
\label{eq:gen_dyn}
\dot{\mathbf{s}}_i = \mathbf{s}_i \times \left[ \nabla_{\mathbf{s}_i} H +
\int_{-\infty}^t \d t' \Gamma(t-t') \dot{s}_{i,z}(t') \hat{\mathbf{z}} + \eta
\hat{\mathbf{z}} \right] 
\end{equation}
where the Keldysh correlator $D^K(t-t') = -2 i
\cexp{\eta_i(t)\eta_i(t')}$ defines the Gaussian noise correlations. Evaluating
this via the fluctuation dissipation relation $D^K = (D^R - D^A)
\coth\left(\frac{\omega}{2T}\right)$ one arrives at
\begin{equation}
\label{eq:eta}
\cexp{\eta_i(t)\eta_j(t')} =  \frac{\delta_{ij}}{4 \pi}
\int_0^\infty \d \omega J(\omega) \coth\left(\frac{\omega}{2 T}\right)
\cos(\omega (t-t')),
\end{equation}
whilst the retarded correlator defines the damping kernel
\begin{equation}
\label{eq:gamma}
\Gamma(t) = \int_{-\infty}^t \d t' D^R(t') = \frac{1}{2 \pi} \int_0^\infty
\frac{\d \omega}{\omega} J(\omega) \cos(\omega t). 
\end{equation}
In Eq~\eqref{eq:gen_dyn} the bare
capacitance has been renormalised by a term generated by the bath in the usual
way\cite{leggett1987}.

Many of the dynamical features of this model can be seen within the
Markovian approximation in which the bath memory is neglected. In this regime
the dynamics are described by the Langevin equation which can be expressed in two
equivalent forms, where $\bm{\eta}_i = \eta_i \hat{\mathbf{z}}$ and
$\mathbf{B}_i = -\nabla_{\mathbf{s}_i} H = \mathbf{h}_{i} + \sum_j
\mathbf{J}_{ij} \mathbf{s}_{j}$ + \ldots, we obtain
\begin{subequations}
\label{eq:zLLG0}
\begin{align}
\label{eq:zLLG}
\dot{\mathbf{s}}_i + \mathbf{s}_i \times \left[\left(\mathbf{B}_i +
\bm{\eta}_i\right) - \gamma \hat{\mathbf{z}}
(\hat{\mathbf{z}} \cdot\dot{\mathbf{s}}_i)\right] &= 0, \\
\label{eq:zLLG2}
\dot{\mathbf{s}}_i + \mathbf{s}_i \times \left(\mathbf{B}_i+\bm{\eta}_i\right) +
\gamma \, \mathbf{s}_i \times \hat{\mathbf{z}} \left[ \hat{\mathbf{z}}
\cdot ( \mathbf{s}_i \times \mathbf{B}_i ) \right] &= 0.
\end{align}
\end{subequations}
The Langevin equation, Eq~(\ref{eq:zLLG}), is a stochastic differential equation
whose solution is given by an ensemble of pure state trajectories. While it is 
possible to obtain the many body dynamics within any manifold of quantum states,
to obtain the classical dynamics of a many qubit system, we have restricted
the pure state trajectories to product states, thus encoding only classical
correlations.

The Langevin equations, Eqs~(\ref{eq:zLLG}) and (\ref{eq:zLLG2}), obtained by this
method include, by construction, the dissipative and stochastic effects induced
by an anisotropic bath. These dynamics are different from those obtained for an
isotropic bath,\footnote{As shown in appendix~\ref{sec:FPE} these differences
also exist at the ensemble level.} Eqs~(\ref{eq:LLG}) and (\ref{eq:LLG2}), with important
consequences.

\section{Dynamics of the Model}
\label{sec:dynamics}
We now explore the novel dynamics in the presence of anisotropic coupling to
the bath in its different parameter regimes. This model, Eqs~(\ref{eq:zLLG})
and (\ref{eq:zLLG2}), describes the dissipative dynamics of a system of
interacting, non-entangled flux qubits, with environmental coupling solely
through the $\hat{s}_z$ operator. There are important differences between the
effects of isotropic environmental couplings, Eqs~(\ref{eq:LLG}) and
(\ref{eq:LLG2}), and anisotropic couplings, Eqs~(\ref{eq:zLLG}) and
(\ref{eq:zLLG2}). The energy conserving dynamics of these two models are
the same, consisting of precession about the external field, and they both relax to the same thermal equilibrium distribution. We shall concentrate upon the case where thermal fluctuations are small in the sense that the equilibrium thermal distribution subtends a small solid angle on the Bloch sphere. This requires that
$\langle \theta^2 \rangle \approx T/B\ll \pi^2$, or alternatively $B \gg T$. In this limit, we may sensibly discuss a large deviation from thermal equilibrium and consider the dissipative relaxation to it. These dynamics are very different in the pressence of an isotropic, Eqs~(\ref{eq:LLG}) and
(\ref{eq:LLG2}), and anisotropic coupling, Eqs~(\ref{eq:zLLG}) and
(\ref{eq:zLLG2}), to the bath.

The Landau-Lifschitz-Gilbert equation constitute a special case where the
Hamiltonian and dissipative dynamics separate in the equations of motion. This
separation is clear when written in polar coordinates. For brevity we neglect stochastic terms and let $\mathbf{B} \parallel \mathbf{z}$, we find
\begin{equation}
\dot{\theta} = - \frac{\gamma s B}{1+\gamma^2 s^2} \sin \theta, \qquad 
\dot{\phi} = \frac{B}{1+\gamma^2 s^2} .
\end{equation}
These equations generate motion in perpendicular directions and, there is no
interplay between their dynamics.

In contrast, in more general dissipative models there is no such separation of
the effects of precession and dissipation, and their interplay remains
important. In the Langevin equation for $\hat{s}_z$ coupling,
eqn.~(\ref{eq:zLLG2}), the system relaxes indirectly,
through the interplay of dynamics and the state dependent modulation of the rate of 
dissipation. The effect of this interplay is highlighted by the appearance of
regions of novel behaviour, visible in fig~\ref{fig:dynamics}, such as
dissipation free precession, retrograde motion, and effective dimensional
reduction.

To discuss the dynamics of the $\hat{s}_z$-coupling model, it is useful to
introduce the timescales $\tau_\mathrm{p}^{-1} = B  = \left|\mathbf{B} \right|$
and $\tau_\mathrm{d}^{-1} = \gamma s B \sin^2 \theta^* $ where $\theta^*$ is the polar angle of the field 
$\mathbf{B}$ from $\mathbf{z}$. $\tau_\mathrm{p}$ and
$\tau_\mathrm{d}$ which are characteristic of the precessional motion and
dissipative motion, respectively. In the limits where these scales are widely
separated, the system's behaviour is dominated by the faster dynamics on short
timescales, whilst some effective dynamics emerge on longer timescales.

\subsection{Weak coupling limit, $\tau_\mathrm{p} \ll  
\tau_\mathrm{d} $}
\label{sec:weakcoupling}
\begin{figure}
\subfloat[][$\theta^* = \tfrac{\pi}{8}, \gamma s = 0.6$]{\label{fig:szweak}
    \includegraphics[width=0.2\textwidth]{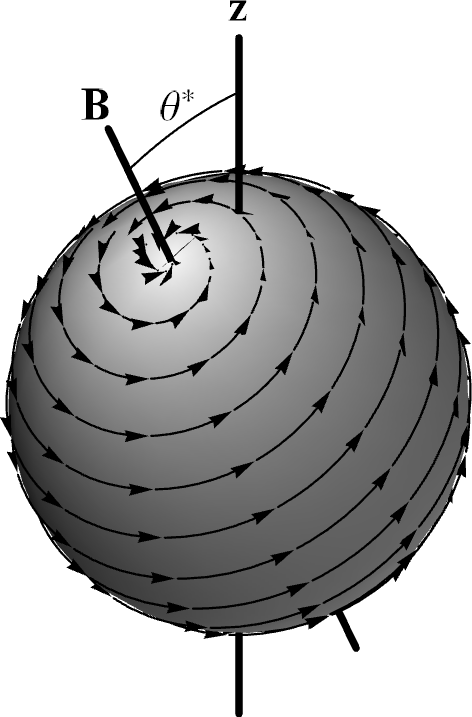}
}
\qquad
\subfloat[][$\theta^* = \tfrac{\pi}{8}, \gamma s = 6 $]{\label{fig:szstrong}
    \includegraphics[width=0.2\textwidth]{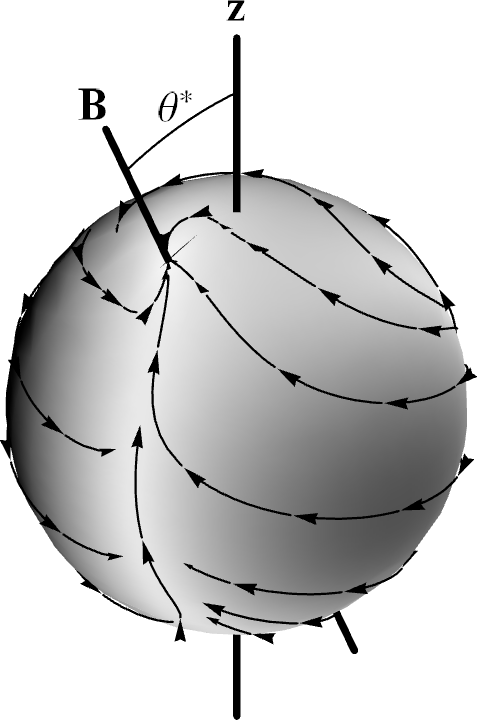}
}
\caption{
{\it Dynamics of spins in the presence of anisotropic dissipation.} The
direction and rate of evolution over the sphere surface are indicated by the streamlines and
grayscale density, darker regions indicate faster evolution.
(a) For {\it weak anisotropic coupling} the state precesses similarly to
isotropic coupling.
(b) For {\it strong anisotropic coupling} dynamics is markedly different. The
system rapidly relaxes to a reduced $\mathrm{O}(2)$ manifold where it undergoes
constrained dynamics. 
}
\label{fig:dynamics}
\end{figure}

For weak coupling, the system's behaviour remains dominated by the precession
found in the fully closed system dynamics. As shown in fig.~\ref{fig:szweak},
the spin will general perform many rotations about the magnetic field,
$\mathbf{B}$,  before reaching the neighbourhood of the ground state. This
allows the dissipation and noise to be averaged over these rotations. In this
regime the Langevin equation may be approximated by a Landau-Lifschitz-Gilbert
equation
\begin{equation}
\label{eq:wLLG}
\mathbf{\dot{s}} + \mathbf{s} \times \left[
\left(\mathbf{B} + \bm{\eta}_\mathrm{eff}\right) + \gamma_\mathrm{eff}
\,\mathbf{s} \times \left( \mathbf{B} + \bm{\eta}_\mathrm{eff} \right)\right]  = 0
\end{equation}
where $\gamma_\mathrm{eff} = \tfrac12 \gamma \sin \theta^*$ is the effective
dissipation, 
This results in dynamics characteristically similar to the isotropic case
(Eqs~(\ref{eq:LLG}) and (\ref{eq:LLG2})), thus the limit $\tau_\mathrm{p} \ll \tau_\mathrm{d} $ offers no novel dynamics.

\subsection{Strong coupling limit, $\tau_\mathrm{p} \gg \tau_\mathrm{d}$}
\label{sec:strongcoupling}
In the limit $\tau_\mathrm{p} \gg \tau_\mathrm{d}$, when the effects of
anisotropy show up most profoundly, the system's dynamics are dominated by the
dissipative term. This term is state dependent and drives rotation about the
$\mathbf{z}$ axis, however the dissipation goes to zero when $\mathbf{B}$,
$\mathbf{s}$ and $\mathbf{z}$ are coplanar.

Exactly how this reveals itself in the dynamics depends upon the bath memory and temperature, though the net effect is similar in all cases: we see a
separation of timescales and a reduction of the full $\mathrm{O}(3)$ qubit
dynamics to effective $\mathrm{O}(2)$ dynamics characterised by fast decay
towards, or oscillation around, a reduced manifold.

\textbf{The Markovian approximation:} In the first instance, it is easiest to
analyse our model within an Markovian approximation. Separating
Eq~\eqref{eq:zLLG2} into the slow $\theta$ and fast $\phi$ dynamics we obtain
\begin{equation}
\begin{aligned}
\dot{\theta} &= - B \sin \theta^* \, \sin \phi, \\
\dot{\phi}&= B \left( \cos \theta^* - \sin \theta^* \left( \cos \phi
\cot \theta + s \gamma \sin \theta \sin \phi \right) \right) + \eta.
\end{aligned}
\label{eq:StochasticProcess}
\end{equation}
where $\phi$ typically relaxes with a
characteristic timescale 
\begin{equation}
\tau_\phi =\frac{1}{ \gamma s B \sin \theta \sin \theta^*}
\label{tauphi}
\end{equation}
 during which time $\theta$ makes a very small change.
Thus we can treat the appearance of $\phi$ in the $\theta$ dynamics as a
stochastic variable sampling the quasi-static distribution of the $\phi$ dynamics.
This distribution is given approximately by $p(\phi) \propto \exp\left[-
A(\theta) \cos (\phi-\phi^*)\right]$ where $A(\theta)^{-1} = \gamma
T \tau_\phi \cos \phi^*$ and $\tan \phi^* = \sin(\theta-\theta^*)/\left(s
\gamma \sin^2 \theta \sin \theta^*\right)$. Equivalent to quasi thermal-equilibrium fluctuations in $\phi$. 
General effective dynamics in the slow $\theta$ variable are obtained in appendix~\ref{sec:SDD}, the high and low $T$ cases are
discussed here:

For $T \ll B$ the system quickly relaxes to a state in which the $\phi$
distribution is sharply peaked around the dynamical fixed point close to $\phi^*
\approx 0$. This confining behaviour, as shown in fig.~\ref{fig:szstrong} occurs when the
dissipative dynamics drive the system towards a one dimensional manifold on the fast
timescale $\tau_\mathrm{d}$, after which a much slower interplay between the
Hamiltonian and dissipative dynamics sees the system relax to its ground state. 
These latter dynamics are described by
\begin{equation}
\label{eq:LowTO2}
\dot{\theta} =  \frac{B \sin \left( \theta^* - \theta \right)}{s \gamma
\sin^2 \theta} + \eta' 
\end{equation}
where the noise $\cexp{\eta(t)}=0$
has correlations $\cexp{\eta'(t)\eta'(t')} = 2 \gamma T (\tau_\phi B \sin
\theta^*)^2 \delta(t-t')$.
In this limit the stochastic effect of the bath is weak (in the sense that thermal fluctuations subtend only a small angle on the Bloch sphere), whereas it strongly biases the trajectories to dissipate energy. In this manner the effect is akin to the trajectory ensemble approach~\cite{garrahan2009first} with a transition to dynamics confined to $\phi \approx \phi^*$. Due to the known relationship between these dyanmics~\cite{kikuchi1991metropolis,nowak2000monte,cheng2006mapping} this Langevin equation can also be related to a Monte Carlo $\mathrm{O}(2)$ model on appropriate timescales.

For $T \gg B$ the noise is sufficiently strong that $\phi$ makes large
excursions away from the dynamical fixed point. However, due to the separation
of timescales, in both cases the long time dynamics are captured by an
effective theory in the slow variable $\theta$. This constitutes a dissipative
reduction of the phase space from O(3) to effective O(2) dynamics. In the $T \gg B$ limit, this reduced dynamics is described by
\begin{equation}
\label{eq:HighTO2}
\dot{\theta} = 
 \frac{B^2 \sin \left(\theta^* - \theta \right) \sin \theta^*}{2 T \gamma
\sin \theta} + \eta' 
\end{equation}
where the noise has correlations $\cexp{\eta'(t)\eta'(t')} = \tau_\phi ( B
\sin \theta^*)^2
\delta(t-t')$.

The limit in which the dynamics become Markovian is subtle. Often Markovian dynamics can be obtained by assuming an Ohmic bath $J(\omega) = 4 \gamma \omega$. When the temperature is much greater than the characteristic frequency of the system (the frequency at which the bath dynamics are sampled) the bath falls in its classical limit, $\coth (\omega /2T) \rightarrow 2T/\omega$, so that Eq.(\ref{eq:eta})  yields a $\delta$-correlated noise. In this regime, however, thermal motion generally dominates--in the sense that thermal fluctuations cover the entire Bloch sphere; $\sqrt{\langle \theta^2 \rangle} \sim \pi$. There are alternative--more physically realistic--bath distributions for which the Markovian limit arises naturally.

\begin{figure}
\subfloat[][]{\label{fig:szdd3d}
    \includegraphics[width=0.2\textwidth]{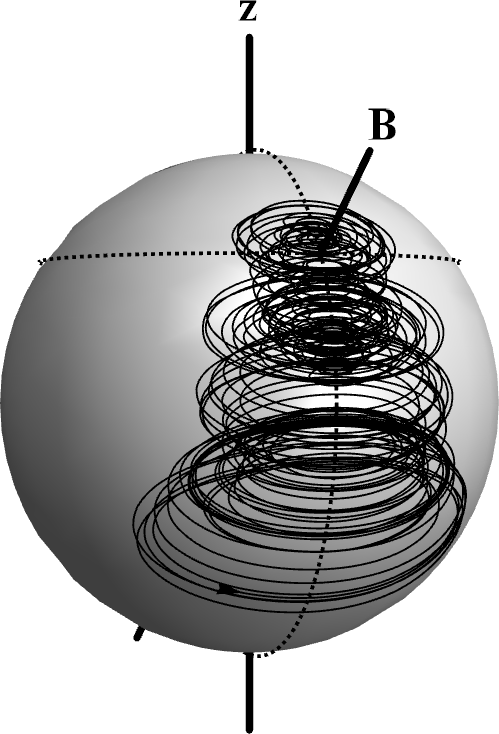}
}
\qquad
\subfloat[][]{\label{fig:szdd2d}
    \includegraphics[width=0.2\textwidth]{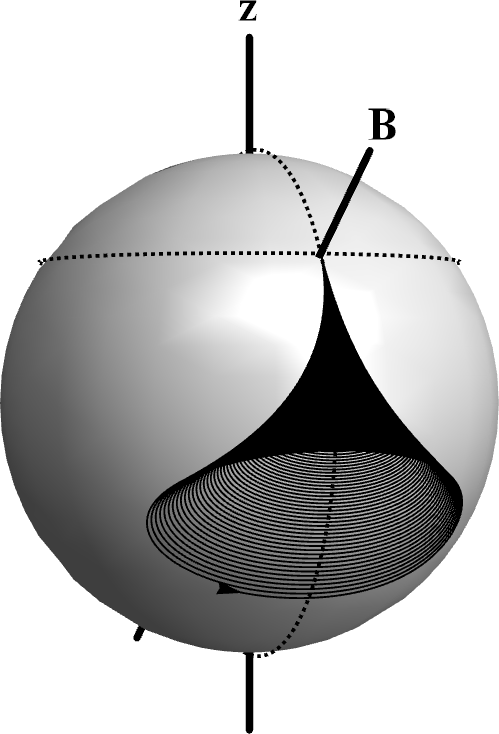}
} 
\caption{
{\it Typical oscillatory stochastic dynamics of spins in the presence of
anisotropic dissipation with a Drude bath.} (a) A trajectory
(solid) with parameters $\gamma = 400$, $B = 50/3 T = 50 \,
\omega_\mathrm{d}$, plotted on the Bloch sphere. (b) The same trajectory
plotted with the unphysical choice of $T=0$ to illustrate the deterministic
part of the dynamics:
The system oscillates around the O(2) manifold with decaying amplitude.
The lines $\phi=0$ and $\theta=\theta^*$ (dashed) are shown, and an arrowhead
indicates the initial state.}
\label{fig:szdd}
\end{figure}

\begin{figure}
\subfloat[][relaxation of $\theta$ up to $t=10^4 / B$ shown by the mean upper bound $\theta_U$, mid-point  $\theta_M$, and lower bound $\theta_L$ of the trajectory envelopes. The the vertical line
indicates the range of plot (b). ]{
\includegraphics[width=240pt,resolution=1000]{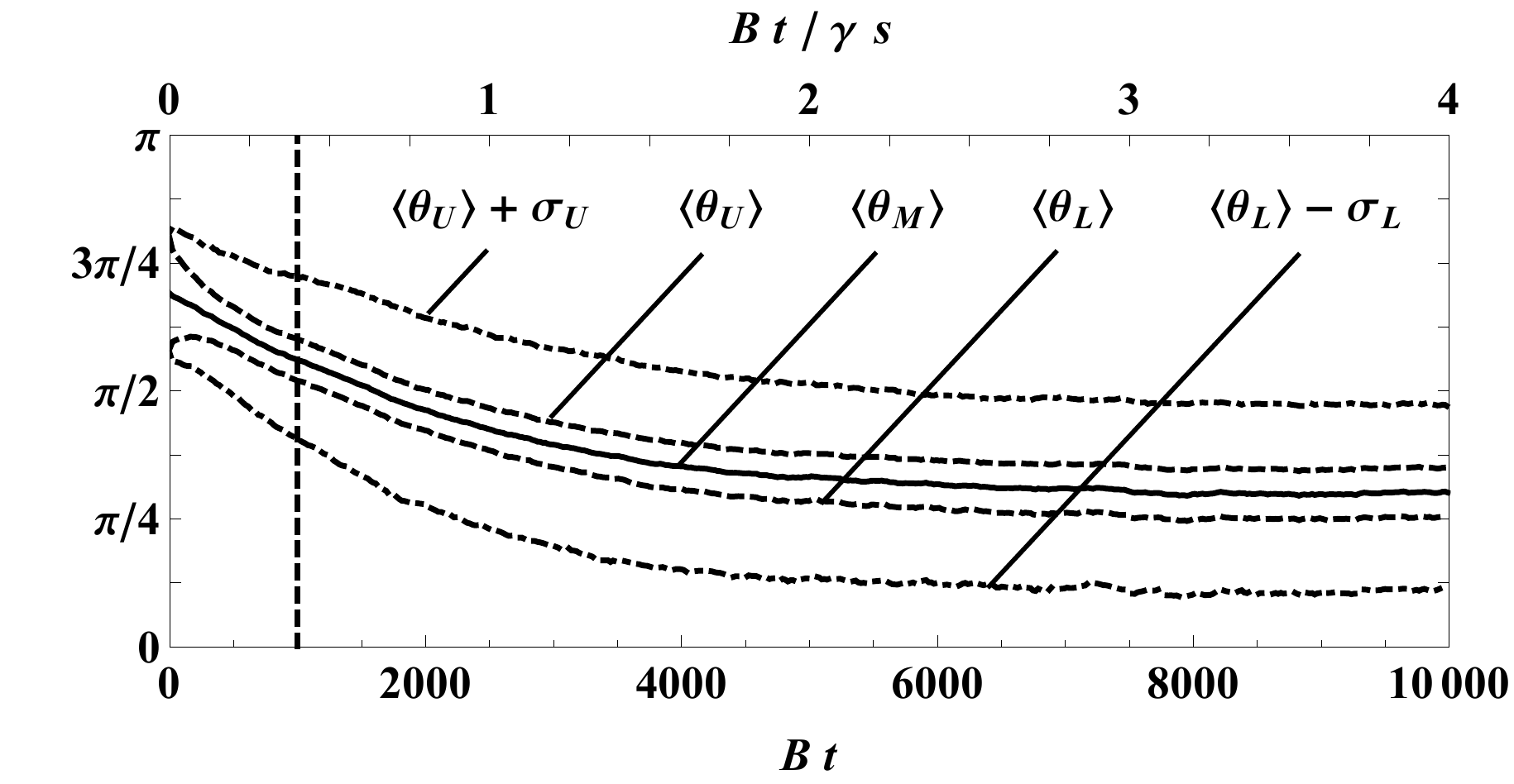}
}

\subfloat[][relaxation of $\phi$ up to $t = 10^3 / B$ shown by the mean upper bound $\phi_U$, mid-point  $\phi_M$, and lower bound $\phi_L$ of the trajectory.
($10\times$ rescaling)]{
\includegraphics[width=240pt,resolution=1000]{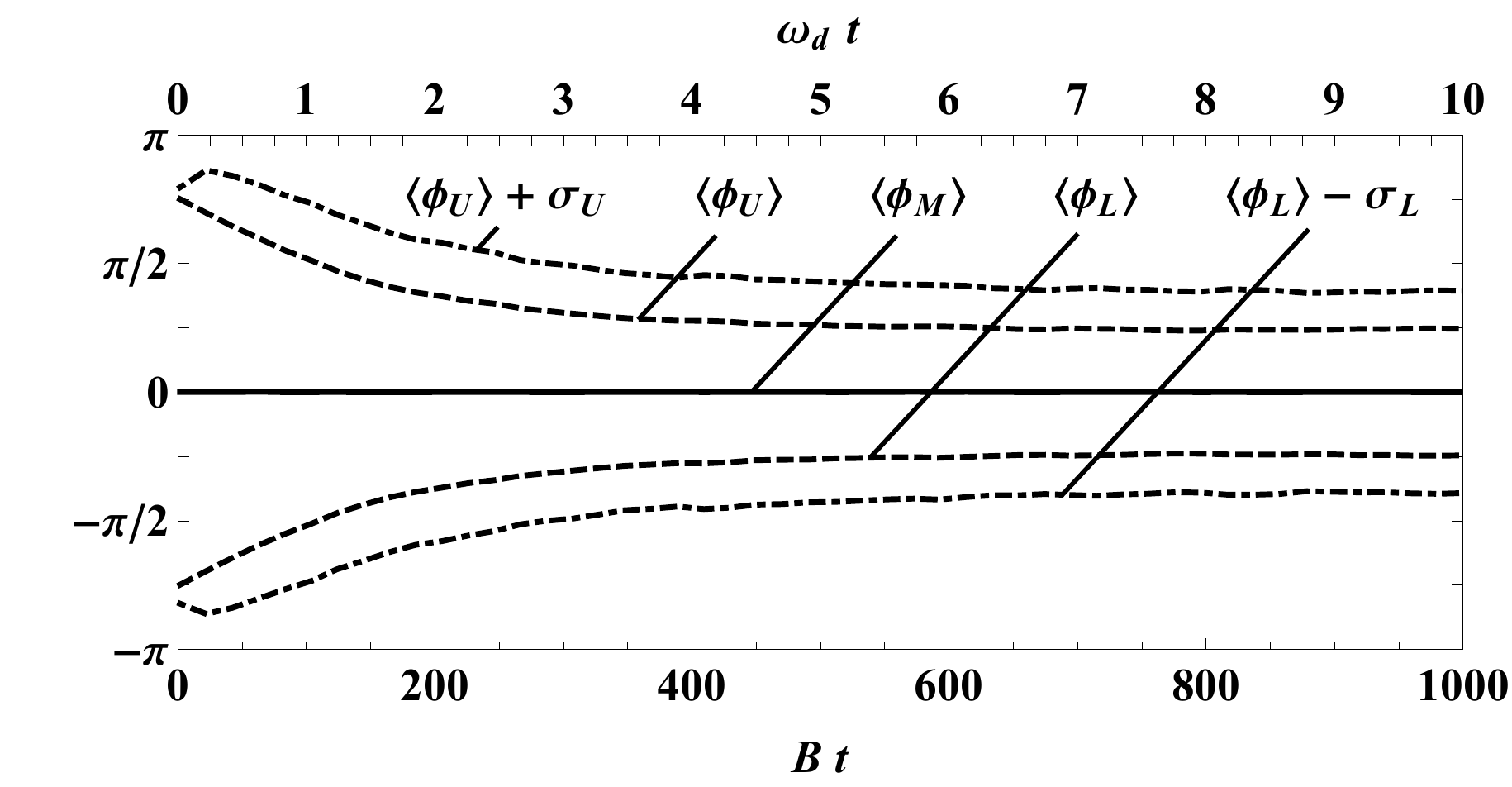}
}

\caption{{\it Different timescales of relaxation in the stochastic dynamics of (a) $\theta$ and (b) $\phi$ when coupled to a Drude bath (Note the different plot ranges).} An ensemble of 1000 spins ($s=1/2$) initially at $\theta = 3 \pi /4$, $\phi = 3 \pi / 4$ when $t = 0$ evolve with a magnetic field in the $\theta^* = \pi/4$, $\phi^* = 0$ direction. The coupling is $\gamma = 5\times10^3$, with energy scales $B = 10 T = 100 \omega_\mathrm{d}$, satisfying $B \gg T \gg \omega_\mathrm{d} \gg B / \gamma s$.On the timescales of $\theta$  dynamics the fast oscillations in the trajectories (see fig~\ref{fig:trajectory}c) means that the trajectories are best characterised by an envelope with upper bound $\theta_U$, lower bound $\theta_L$ and mid-point $\theta_M$, these are simply read off the oscillating trajectory as shown in  fig~\ref{fig:trajectory}c. Since the initial conditions is an extrema of the fast oscillations the initial point lies on $\theta_U$ and $\phi_U$. The ensemble averages $\cexp{\theta_M}$ (solid) and $\cexp{\theta_U}, \cexp{\theta_L}$ (dashed) are shown with $\cexp{\theta_U} + \sigma_U = \cexp{\theta_U}+\sqrt{\cexp{\theta_U^2}-\cexp{\theta_U}^2}$, and $\cexp{\theta_L} - \sigma_L = \cexp{\theta_L}-\sqrt{\cexp{\theta_L^2}-\cexp{\theta_L}^2}$ (both dot--dashed) illustrating the ensemble width. (a) The slow theta coordinate relaxes towards the equilibrium value $\theta^* = \pi/4$ on a timescale $ \gamma s / B$, approaching it at $t \approx 10^4/B$. The vertical line indicates the range of plot (b). (b) The same statistics are presented for the $\phi$ dynamics: The $\phi$ dynamics relaxes to its equilibrium distribution much faster on a characteristic timescale $\tau \sim 1/\omega_\mathrm{d}$ and is fully relaxed by $t \approx 500/B$.}
\label{fig:ensemble}
\end{figure}

\begin{figure}

\subfloat[][$\theta$ relaxation on a long timescale, up to $t = 10^4/B$.]{
\includegraphics[width=240pt,resolution=1000]{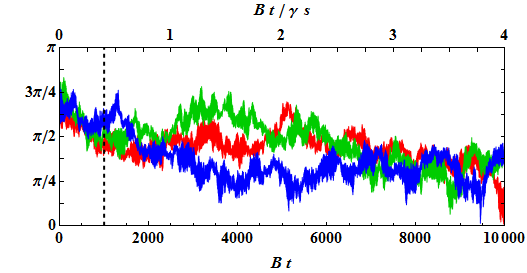}
}

\subfloat[][$\phi$ relaxation on a shorter timescale, plotted up to $t = 10^3/B$ ($10 \times$ rescaling).]{
\includegraphics[width=240pt,resolution=1000]{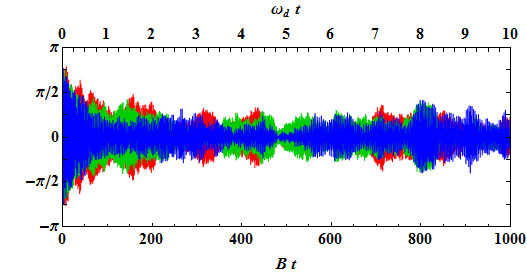}
} 

\subfloat[][very short timescale oscillatory behaviour in $\theta$ induced by bath memory (see eq~\ref{eq:DrudeNoise}).]{
\includegraphics[width=240pt,resolution=1000]{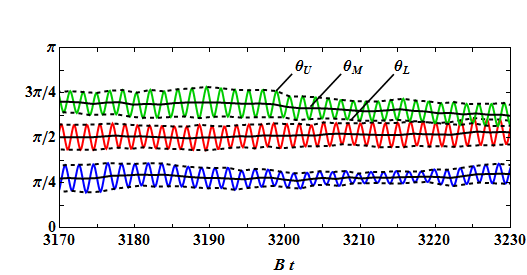}
}  

\caption{{\it Different dynamical timescales of typical trajectories in $\theta$ and $\phi$ when coupled to a Drude bath.} A plot of three sample trajectories from the ensemble studied in fig~\ref{fig:ensemble}. Being drawn from the ensemble these spins ($s=1/2$) evolved in the same conditions: initially prepared at $\theta = 3 \pi /4$, $\phi = 3 \pi / 4$ at $t = 0$ and evolve with a magnetic field in the $\theta = \pi/4$, $\phi = 0$ direction. The coupling is $\gamma = 5\times10^3$, with energy scales $B = 10 T = 100 \omega_\mathrm{d}$, satisfying $B \gg T \gg \omega_\mathrm{d} \gg B / \gamma s$. (a) Trajectories in $\theta$ relax on a a long time scale. The vertical dashed line indicates the range of plot (b). (b) $\phi$ relaxes on a shorter timescale, the confinement of $\phi$ is evidenced by the typically small excursions from $\phi = 0$. (c) Oscillatory behaviour induced by the bath occurring on shorter timescales $\tau \sim \sqrt{\tau_\phi/\omega_\mathrm{d}}$ is plotted for $\theta$, similar behaviour occurs for $\phi$. This behaviour is expanded on in appendix~\ref{sec:DD}, where the oscillations appear in eq~\ref{eq:DrudeNoise}. Each oscillatory trajectory can be characterised by the upper and lower edges $\theta_U$ and $\theta_L$ of its envelope, and its midpoint $\theta_M$. The ensemble statistics of these quantities are studied in fig~\ref{fig:ensemble}.} 
\label{fig:trajectory}
\end{figure}

\textbf{Drude dynamics:} 
The Drude bath has density of states $J(\omega)
= 4 \gamma \omega \frac{ \omega_\mathrm{d}^2}{\omega^2+\omega_\mathrm{d}^2}$, where the
Drude frequency, $\omega_\mathrm{d}$, defines the bath cutoff. The new energy scale allows the bath to be in its classical limit, $T\gg \omega_\mathrm{d}$, even when the equilibrium thermal fluctuations of the system are small, $T \ll B$. It is then sensible to discuss the dissipation-dominated realaxation to this equilibrium distribution. 

The Drude bath has a memory on timescales determined by the cut-off,
$\omega_d^{-1}$.  The dynamical dependence on the history of the bath is
captured by introducing a time-dependent field term
\begin{equation}
\label{eq:DLLG}
\begin{aligned}
\dot{\mathbf{s}}_i &= -\mathbf{s}_i \times \left(\mathbf{B}_i +
\mathbf{B}_\mathrm{diss} \right),
\\
\dot{\mathbf{B}}_\mathrm{diss} &= 
 -\omega_\mathrm{d} \left( \mathbf{B}_\mathrm{diss} - \gamma \hat{\mathbf{z}}
 \left[ \hat{\mathbf{z}} \cdot \left( \mathbf{s}_i \times \mathbf{B}_i  \right)
 \right] - \bm{\eta} \right),
\end{aligned}
\end{equation}
where $\cexp{\eta_z(t)\eta_z(t'))} = 2 \gamma T \delta(t-t')$ and as before and
we take $ \omega_d \ll T$. This is shown in detail in Appendix~\ref{sec:DD}.
The main effect of the bath is to induce oscillations in $\phi$ about the
Markovian trajectory. In this strongly
dissipative limit the oscillations have a frequency
$\sqrt{\omega_\mathrm{d}/\tau_\phi}$ and describe 
fluctuations that are small in the $\theta$
direction and decay away on the bath memory timescale $1/\omega_\mathrm{d}$.
When the oscillation decay is much faster than the decay of $\theta$---i.e
$\omega_\mathrm{d} \gg B /s \gamma$---we recover the Markovian O(2)
dynamics described by eqs~\eqref{eq:LowTO2}. A derivation of this is shown in
appendix~\ref{sec:DD}.
The characteristic qualities of a typical trajectory are depicted in
figure~\ref{fig:szdd}, whilst a simulation of the dynamics of
equation~\eqref{eq:DLLG} showing the separation of characteristic timescales is
shown in figures~\ref{fig:ensemble} and~\ref{fig:trajectory}.

To conclude this discussion we summarise the physical limits we have studied. We have assumed that the equilibrium thermal fluctuations are small $B \gg T$ and that the environmental coupling is strong. We have further assumed that temperature prevents quantum correlations
persisting in the bath, $T\gg \omega_\mathrm{d}$--though this condition may
be relaxed.
Our analysis shows an effective reduction in phase space from,
$\mathrm{O}(3)$ to $\mathrm{O}(2)$, resulting from the confinement of the
$\phi$ coordinate to a typically small region around $\phi = 0$.
This occurs when $B \gg T \gg \omega_\mathrm{d} \gg B / \gamma s$, for both
Markovian and Drude dynamics. In this parameter regime we have also shown that
when coupled to the more general Drude bath, the system is described by the Markovian $\mathrm{O}(2)$ dynamics of equation~\eqref{eq:LowTO2} on timescales greater than the bath memory, $1/\omega_\mathrm{d}$.

\section{Consequences of Anisotropic Dissipation}
The Langevin equation derived in the previous section exhibits markedly
distinct behaviours in different regimes. This has implications both for attempts to fit experimental data to such models and, ultimately, for the usefulness of a system described by them for computation.

\subsection{Different behaviours for the same system} 
\label{sec:DifferentBehaviour}
For a qubit to be useful for quantum computation it must be sufficiently manipulable. For the simplest system of a single qubit, this in effect requires the implementation of at least two non-parallel magnetic fields. The field, $\mathbf{B}$, is thus assumed to be a tunable parameter of the system, whereas the environmental coupling, $\gamma$, is fixed at some finite value.

For the anisotropic coupling to the environment studied here, the strength of dissipation depends not only upon the fixed parameter $\gamma$, but also upon the orientation of the magnetic field, $\bf {B}$.
This is important when characterising such a device. In particular a system that is analysed under conditions when
$\mathbf{B}$ and $\mathbf{z}$ are nearly aligned will appear weakly
coupled, whereas for other orientations of $\mathbf{B}$ the
dynamics may be entirely dominated by environment-induced dissipative dynamics.

It is a general feature of qubit systems that inhomogeneous environmental
couplings will result in dynamics that are correspondingly
inhomogeneous~\cite{weiss1999,dube2001,prokof2000,leggett1987}.
We see here that certain states evolve with minimal dissipation, whilst others are dominated by dissipative or noisy dynamics. In
extreme cases this may amount to a reduction in the effective state space.

\subsection{A model for lossy qubit arrays}

A natural application of our analysis is to understand some puzzling features of the D-wave machine~\cite{johnson2011quantum}. This machine consists of a tuneable array of coupled Josephson junctions whose dynamics may be controlled to perform a quantum annealing or adiabatic computation. Various models have been posited for this system, including Bloch-Redfield simulations, Landau-Lifshitz-Gilbert models and two different $\mathrm{O}(2)$ models (a thorough review can be found in Ref~\onlinecite{vinci2014}). 
Our analysis sheds light upon the relationship between these different models and the fact that apparently quite different models yield surprisingly similar results. In particular, the salient features of the two $\mathrm{O}(2)$ models proposed in the under\cite{smolin2013}- and over\cite{shin2014,shinagain2014}-damped regimes can be found in different limits of the Landau-Lifshitz-Gilbert model when proper attention is paid to the effects of the bath. These models are appropriate in a limit where decoherence renders entanglement effects negligible. Bloch-Redfield simulations --- the only one of the above to include entanglement --- are expected to yield similar results in this limit. 

In the {\it underdamped limit}, the qubit precesses about the adiabatic minimum.
Projecting this motion onto the polar angle results in harmonic oscillations of the polar angle about its adiabatic minimum. This is essentially the  $\mathrm{O}(2)$ model of Ref~\onlinecite{smolin2013}, though strictly the effective kinetic term depends upon the local effective field felt by the quantum bit and so varys through the computation\footnote{It can be derived in a spinwave-like expansion, integrating out the components that do not lie in the great circle traversed by the adiabatic minimum.}. This model is expected to accurately reflect the dynamics up to the point where it deviates markedly from adiabatic, accounting for its success in predicting the probability of correctly performing  adiabatic computation in some circumstances~\cite{boixo2013}.

In this manuscript we have focussed upon the {\it overdamped limit}. A model of over-damped classical rotors undergoing a thermal exploration of the $\mathrm{O}(2)$ state space was introduced  in Ref~\onlinecite{shin2014}. This model reproduced 
additional, apparently quantum, effects~\cite{boixo2014}, though it is interesting that the underdamped non-thermal $\mathrm{O}(2)$ model was also able to reproduce these results. Further statistics presented in evidence of quantum effects~\cite{vinci2014} were reproduced only by the over-damped classical model~\cite{shinagain2014}---and not the underdamped model. The overdamped model of Refs~\onlinecite{shin2014} and~\onlinecite{shinagain2014} can be obtained by artificially confining the motion of overdamped classical $\mathrm{O}(3)$ spins to the $\mathrm{O}(2)$ phase space~\cite{albash2014}. Our analysis in section~\ref{sec:strongcoupling} shows that, remarkably,  anisotropic dissipation can bias the dynamics towards just such a confinement: the $\mathrm{O}(3)$ dynamics of dissipative spin dynamics (Eqs.~\eqref{eq:zLLG0} and~\eqref{eq:DLLG}) are reduced to the effective $\mathrm{O}(2)$ dynamics of Eq.~\eqref{eq:LowTO2} as illustrated in Figs.~\ref{fig:ensemble} and \ref{fig:trajectory}. 
An unanticipated feature of this relaxation to an $O(2)$ manifold is the restriction of the system dynamics  to half of the $O(2)$ submanifold. The whole of this submanifold is a fixed point of the initial rapid decay, but half forms a stable and half an unstable fixed point. 

The representation of over-damped classical rotors in Refs~\onlinecite{shin2014} and~\onlinecite{shinagain2014} is at first glance rather different from the Langevin equation of Eq.\eqref{eq:LowTO2}. However, Metropolis-Hastings dynamics describes a dissipative relaxation to thermal equilibrium and have been used previously to model dynamics described by a Langevin equation~\cite{kikuchi1991metropolis,nowak2000monte,cheng2006mapping}\textsuperscript{,}\footnote{The relation of the Metropolis algorithm to Fokker-Planck diffusion has been noted previously in general terms in Ref~\onlinecite{kikuchi1991metropolis} whilst other Monte--Carlo methods have been specifically connected with Langevin type dynamics in
Refs~\onlinecite{nowak2000monte} and~\onlinecite{cheng2006mapping}}. The {\it ad hoc.} model of Refs~\onlinecite{shin2014} and~\onlinecite{shinagain2014} do not include the biasing of trajectories to half of the $O(2)$ manifold that naturally arises from microscopic considerations. However, the dynamics of the component of qubit projected onto the direction of the local field is rather similar (see Appendix~\ref{sec:FPE}), perhaps accounting for the success of Refs~\onlinecite{shin2014} and~\onlinecite{shinagain2014} despite their models not being strictly derivable from microscopic considerations. 

These analyses raise an immediate question of whether the D-wave system is in an {\it overdamped or underdamped limit}. This is subtle. As discussed in \ref{sec:DifferentBehaviour}, the strength of damping depends upon the microscopic details of coupling to the bath, the orientation of the effective field relative to the $z$-axis, and the instantaneous position of the qubit on the Bloch sphere. Because of the latter effects, the dissipation is largest at the start of D-wave computation, when the effective field and Bloch spins are in the $xy$-plane, and decreases to zero as the computation proceeds. Even when the coupling to the bath is strong, and the initial dynamics overdamped, there is a transition to underdamped dynamics at some point in the computation. Whether or not the over- or under-damped dynamics determines the success or failure of a computation depends upon precisely when this cross-over occurs.

So is the D-wave system initially over-or under-damped?
This question does not appear to be resolved by published experimental data. 
To our knowledge no direct measurements of $T_1$ or $T_2$ times on the D-Wave are available. 
Measurements of high frequency flux noise {\it via} macroscopic resonant tunnelling~\cite{lanting2011probing} indicate that above a cutoff $\omega_{\mathrm{HF}} = 0.5\,\mathrm{GHz}$ noise is  Ohmic with $\gamma_{\mathrm{HF}} = \Phi_0^2 S_{\Phi}(0)/8 \hbar k_\mathrm{B} T L^2 \approx 0.5$, where the values for inductance $L = 265.4\text{pH}$ and shunt resistance $R = 2 k T L^2/S_{\Phi}(0) = 20k\Omega$ measured in Ref~\onlinecite{lanting2011probing}, and $\Phi_0$ is the magnetic flux quantum. However, our analysis shows that the qubit is also sensitive to noise with frequency lower than the system frequency. At the lowest frequencies, the noise is of $1/f$ form~\cite{harris2010experimental}. There is a large window between the high frequency\cite{lanting2011probing} and the low frequency measurements\cite{harris2010experimental} in which the noise has not been directly measured. The measured current noise characteristics do not, therefore,  preclude the possibility of $\gamma >1$ and initial over-damped dynamics.

\section{Conclusion}
Our main result has been to show how anisotropic dissipation can bias quantum trajectories towards particular sub-manifolds of the system's Hilbert space. We have found a Langevin description of the dynamics of qubits that allows for anisotropic coupling to the environment. This is a natural generalisation of the Landau-Lifschitz-Gilbert equations which describe the dissipative dynamics of spins with isotropic coupling to the environment. The fluctuation-dissipation relation has the important consequence that the anisotropic noise generated by this coupling inevitably leads to anisotropic dissipation.

This model applies explicitly to qubits experiencing dissipation due to fluctuations in the level separation (environmental coupling to the $\hat{s}_z$ operator). When the coupling to the bath is strong the anisotropic dissipation drives rapid relaxation to a reduced $\mathrm{O}(2)$ manifold of constrained dynamics. 
This emergence of this effective dynamics from the underlying microscopics reproduces some salient features of the 
dynamics of the models of Refs~\onlinecite{shin2014,shinagain2014}.
These models were capable of reproducing several observed behaviours of the D-Wave machine previously believed to evidence quantum dynamics. This highlights the necessity of understanding the dynamics in dissipative and strong coupling cases when interpreting the dynamics of an experimental system.

Entanglement, which we neglect here, is crucial for full quantum dynamics, and
necessary to get the exponential scaling between the quantum state space and
number of qubits. It has been argued recently in
Refs~\onlinecite{crowley2014,bauer2015} to act as a resource for adiabatic
computation. These works modelled quantum adiabatic computation with artificial
constraints on the entanglement analogous to the artificial constraint of a
local subsystem to an $\mathrm{O}(2)$ manifold.
However one may anticipate that the effects of dissipation could
naturally bias the trajectories to these restricted manifolds. Understanding
this will be key to determining how best to use limited or dissipating
entanglement resources in computation.

Understanding of the effects of state-dependent noise and anisotropic coupling
to the environment is crucial for the proper control of quantum devices. As we
have shown in the case of the D-Wave machine, these effects can bias the system
dynamics in unexpected ways. Used constructively, this may be harnessed to
useful ends. If ignored, the dynamics may completely different from that
intended.

\section{Acknowledgements} 

This research was supported by the EPSRC under grants EP/K02163X/1 and EP/I004831/2.

\bibliography{bibliography}

\begin{thebibliography}{47}%
\makeatletter
\providecommand \@ifxundefined [1]{%
 \@ifx{#1\undefined}
}%
\providecommand \@ifnum [1]{%
 \ifnum #1\expandafter \@firstoftwo
 \else \expandafter \@secondoftwo
 \fi
}%
\providecommand \@ifx [1]{%
 \ifx #1\expandafter \@firstoftwo
 \else \expandafter \@secondoftwo
 \fi
}%
\providecommand \natexlab [1]{#1}%
\providecommand \enquote  [1]{``#1''}%
\providecommand \bibnamefont  [1]{#1}%
\providecommand \bibfnamefont [1]{#1}%
\providecommand \citenamefont [1]{#1}%
\providecommand \href@noop [0]{\@secondoftwo}%
\providecommand \href [0]{\begingroup \@sanitize@url \@href}%
\providecommand \@href[1]{\@@startlink{#1}\@@href}%
\providecommand \@@href[1]{\endgroup#1\@@endlink}%
\providecommand \@sanitize@url [0]{\catcode `\\12\catcode `\$12\catcode
  `\&12\catcode `\#12\catcode `\^12\catcode `\_12\catcode `\%12\relax}%
\providecommand \@@startlink[1]{}%
\providecommand \@@endlink[0]{}%
\providecommand \url  [0]{\begingroup\@sanitize@url \@url }%
\providecommand \@url [1]{\endgroup\@href {#1}{\urlprefix }}%
\providecommand \urlprefix  [0]{URL }%
\providecommand \Eprint [0]{\href }%
\providecommand \doibase [0]{http://dx.doi.org/}%
\providecommand \selectlanguage [0]{\@gobble}%
\providecommand \bibinfo  [0]{\@secondoftwo}%
\providecommand \bibfield  [0]{\@secondoftwo}%
\providecommand \translation [1]{[#1]}%
\providecommand \BibitemOpen [0]{}%
\providecommand \bibitemStop [0]{}%
\providecommand \bibitemNoStop [0]{.\EOS\space}%
\providecommand \EOS [0]{\spacefactor3000\relax}%
\providecommand \BibitemShut  [1]{\csname bibitem#1\endcsname}%
\let\auto@bib@innerbib\@empty
\bibitem [{\citenamefont {Landau}\ and\ \citenamefont
  {Lifshitz}(1935)}]{landau1935}%
  \BibitemOpen
  \bibfield  {author} {\bibinfo {author} {\bibfnamefont {L.~D.}\ \bibnamefont
  {Landau}}\ and\ \bibinfo {author} {\bibfnamefont {E.}~\bibnamefont
  {Lifshitz}},\ }\href@noop {} {\bibfield  {journal} {\bibinfo  {journal}
  {Phys. Z. Sowjetunion}\ }\textbf {\bibinfo {volume} {8}},\ \bibinfo {pages}
  {101} (\bibinfo {year} {1935})}\BibitemShut {NoStop}%
\bibitem [{\citenamefont {Gilbert}(2004)}]{gilbert2004}%
  \BibitemOpen
  \bibfield  {author} {\bibinfo {author} {\bibfnamefont {T.~L.}\ \bibnamefont
  {Gilbert}},\ }\href@noop {} {\bibfield  {journal} {\bibinfo  {journal}
  {Magnetics, IEEE Transactions on}\ }\textbf {\bibinfo {volume} {40}},\
  \bibinfo {pages} {3443} (\bibinfo {year} {2004})}\BibitemShut {NoStop}%
\bibitem [{\citenamefont {Leggett}\ \emph {et~al.}(1987)\citenamefont
  {Leggett}, \citenamefont {Chakravarty}, \citenamefont {Dorsey}, \citenamefont
  {Fisher}, \citenamefont {Garg},\ and\ \citenamefont {Zwerger}}]{leggett1987}%
  \BibitemOpen
  \bibfield  {author} {\bibinfo {author} {\bibfnamefont {A.~J.}\ \bibnamefont
  {Leggett}}, \bibinfo {author} {\bibfnamefont {S.}~\bibnamefont
  {Chakravarty}}, \bibinfo {author} {\bibfnamefont {A.~T.}\ \bibnamefont
  {Dorsey}}, \bibinfo {author} {\bibfnamefont {M.~P.~A.}\ \bibnamefont
  {Fisher}}, \bibinfo {author} {\bibfnamefont {A.}~\bibnamefont {Garg}}, \ and\
  \bibinfo {author} {\bibfnamefont {W.}~\bibnamefont {Zwerger}},\ }\href
  {\doibase 10.1103/RevModPhys.59.1} {\bibfield  {journal} {\bibinfo  {journal}
  {Rev. Mod. Phys.}\ }\textbf {\bibinfo {volume} {59}},\ \bibinfo {pages} {1}
  (\bibinfo {year} {1987})}\BibitemShut {NoStop}%
\bibitem [{\citenamefont {Vinci}\ \emph {et~al.}(2014)\citenamefont {Vinci},
  \citenamefont {Albash}, \citenamefont {Mishra}, \citenamefont {Warburton},\
  and\ \citenamefont {Lidar}}]{vinci2014}%
  \BibitemOpen
  \bibfield  {author} {\bibinfo {author} {\bibfnamefont {W.}~\bibnamefont
  {Vinci}}, \bibinfo {author} {\bibfnamefont {T.}~\bibnamefont {Albash}},
  \bibinfo {author} {\bibfnamefont {A.}~\bibnamefont {Mishra}}, \bibinfo
  {author} {\bibfnamefont {P.~A.}\ \bibnamefont {Warburton}}, \ and\ \bibinfo
  {author} {\bibfnamefont {D.~A.}\ \bibnamefont {Lidar}},\ }\href@noop {}
  {\bibfield  {journal} {\bibinfo  {journal} {arXiv preprint arXiv:1403.4228}\
  } (\bibinfo {year} {2014})}\BibitemShut {NoStop}%
\bibitem [{\citenamefont {Smolin}\ and\ \citenamefont
  {Smith}(2014)}]{smolin2013}%
  \BibitemOpen
  \bibfield  {author} {\bibinfo {author} {\bibfnamefont {J.~A.}\ \bibnamefont
  {Smolin}}\ and\ \bibinfo {author} {\bibfnamefont {G.}~\bibnamefont {Smith}},\
  }\href {\doibase 10.3389/fphy.2014.00052} {\bibfield  {journal} {\bibinfo
  {journal} {Frontiers in Physics}\ }\textbf {\bibinfo {volume} {2}} (\bibinfo
  {year} {2014}),\ 10.3389/fphy.2014.00052}\BibitemShut {NoStop}%
\bibitem [{\citenamefont {Wang}\ \emph {et~al.}(2013)\citenamefont {Wang},
  \citenamefont {R{\o}nnow}, \citenamefont {Boixo}, \citenamefont {Isakov},
  \citenamefont {Wang}, \citenamefont {Wecker}, \citenamefont {Lidar},
  \citenamefont {Martinis},\ and\ \citenamefont {Troyer}}]{wang2013}%
  \BibitemOpen
  \bibfield  {author} {\bibinfo {author} {\bibfnamefont {L.}~\bibnamefont
  {Wang}}, \bibinfo {author} {\bibfnamefont {T.~F.}\ \bibnamefont {R{\o}nnow}},
  \bibinfo {author} {\bibfnamefont {S.}~\bibnamefont {Boixo}}, \bibinfo
  {author} {\bibfnamefont {S.~V.}\ \bibnamefont {Isakov}}, \bibinfo {author}
  {\bibfnamefont {Z.}~\bibnamefont {Wang}}, \bibinfo {author} {\bibfnamefont
  {D.}~\bibnamefont {Wecker}}, \bibinfo {author} {\bibfnamefont {D.~A.}\
  \bibnamefont {Lidar}}, \bibinfo {author} {\bibfnamefont {J.~M.}\ \bibnamefont
  {Martinis}}, \ and\ \bibinfo {author} {\bibfnamefont {M.}~\bibnamefont
  {Troyer}},\ }\href@noop {} {\bibfield  {journal} {\bibinfo  {journal} {arXiv
  preprint arXiv:1305.5837}\ } (\bibinfo {year} {2013})}\BibitemShut {NoStop}%
\bibitem [{\citenamefont {Shin}\ \emph
  {et~al.}(2014{\natexlab{a}})\citenamefont {Shin}, \citenamefont {Smith},
  \citenamefont {Smolin},\ and\ \citenamefont {Vazirani}}]{shin2014}%
  \BibitemOpen
  \bibfield  {author} {\bibinfo {author} {\bibfnamefont {S.~W.}\ \bibnamefont
  {Shin}}, \bibinfo {author} {\bibfnamefont {G.}~\bibnamefont {Smith}},
  \bibinfo {author} {\bibfnamefont {J.~A.}\ \bibnamefont {Smolin}}, \ and\
  \bibinfo {author} {\bibfnamefont {U.}~\bibnamefont {Vazirani}},\ }\href@noop
  {} {\bibfield  {journal} {\bibinfo  {journal} {arXiv preprint
  arXiv:1401.7087}\ } (\bibinfo {year} {2014}{\natexlab{a}})}\BibitemShut
  {NoStop}%
\bibitem [{\citenamefont {Kamenev}(2011)}]{kamenev2011}%
  \BibitemOpen
  \bibfield  {author} {\bibinfo {author} {\bibfnamefont {A.}~\bibnamefont
  {Kamenev}},\ }\href {https://books.google.co.uk/books?id=CwlrUepnla4C} {\emph
  {\bibinfo {title} {Field Theory of Non-Equilibrium Systems}}}\ (\bibinfo
  {publisher} {Cambridge University Press},\ \bibinfo {year}
  {2011})\BibitemShut {NoStop}%
\bibitem [{\citenamefont {Orth}\ \emph {et~al.}(2010)\citenamefont {Orth},
  \citenamefont {Imambekov},\ and\ \citenamefont {Le~Hur}}]{orth2010}%
  \BibitemOpen
  \bibfield  {author} {\bibinfo {author} {\bibfnamefont {P.~P.}\ \bibnamefont
  {Orth}}, \bibinfo {author} {\bibfnamefont {A.}~\bibnamefont {Imambekov}}, \
  and\ \bibinfo {author} {\bibfnamefont {K.}~\bibnamefont {Le~Hur}},\ }\href
  {\doibase 10.1103/PhysRevA.82.032118} {\bibfield  {journal} {\bibinfo
  {journal} {Phys. Rev. A}\ }\textbf {\bibinfo {volume} {82}},\ \bibinfo
  {pages} {032118} (\bibinfo {year} {2010})}\BibitemShut {NoStop}%
\bibitem [{\citenamefont {Green}(2006)}]{green2006}%
  \BibitemOpen
  \bibfield  {author} {\bibinfo {author} {\bibfnamefont {A.~G.}\ \bibnamefont
  {Green}},\ }\href {\doibase 10.1103/PhysRevB.73.140506} {\bibfield  {journal}
  {\bibinfo  {journal} {Phys. Rev. B}\ }\textbf {\bibinfo {volume} {73}},\
  \bibinfo {pages} {140506} (\bibinfo {year} {2006})}\BibitemShut {NoStop}%
\bibitem [{\citenamefont {Schmid}(1982)}]{schmid1982}%
  \BibitemOpen
  \bibfield  {author} {\bibinfo {author} {\bibfnamefont {A.}~\bibnamefont
  {Schmid}},\ }\href@noop {} {\bibfield  {journal} {\bibinfo  {journal}
  {Journal of Low Temperature Physics}\ }\textbf {\bibinfo {volume} {49}},\
  \bibinfo {pages} {609} (\bibinfo {year} {1982})}\BibitemShut {NoStop}%
\bibitem [{\citenamefont {Weiss}(1999)}]{weiss1999}%
  \BibitemOpen
  \bibfield  {author} {\bibinfo {author} {\bibfnamefont {U.}~\bibnamefont
  {Weiss}},\ }\href@noop {} {\emph {\bibinfo {title} {Quantum dissipative
  systems}}},\ Vol.~\bibinfo {volume} {10}\ (\bibinfo  {publisher} {World
  Scientific},\ \bibinfo {year} {1999})\BibitemShut {NoStop}%
\bibitem [{\citenamefont {Kubo}\ and\ \citenamefont
  {Hashitsume}(1970)}]{Kubo1970}%
  \BibitemOpen
  \bibfield  {author} {\bibinfo {author} {\bibfnamefont {R.}~\bibnamefont
  {Kubo}}\ and\ \bibinfo {author} {\bibfnamefont {N.}~\bibnamefont
  {Hashitsume}},\ }\href {\doibase 10.1143/PTPS.46.210} {\bibfield  {journal}
  {\bibinfo  {journal} {Progress of Theoretical Physics Supplement}\ }\textbf
  {\bibinfo {volume} {46}},\ \bibinfo {pages} {210} (\bibinfo {year} {1970})},\
  \Eprint
  {http://arxiv.org/abs/http://ptps.oxfordjournals.org/content/46/210.full.pdf+html}
  {http://ptps.oxfordjournals.org/content/46/210.full.pdf+html} \BibitemShut
  {NoStop}%
\bibitem [{\citenamefont {Hohenberg}\ and\ \citenamefont
  {Halperin}(1977)}]{hohenberg1977}%
  \BibitemOpen
  \bibfield  {author} {\bibinfo {author} {\bibfnamefont {P.~C.}\ \bibnamefont
  {Hohenberg}}\ and\ \bibinfo {author} {\bibfnamefont {B.~I.}\ \bibnamefont
  {Halperin}},\ }\href {\doibase 10.1103/RevModPhys.49.435} {\bibfield
  {journal} {\bibinfo  {journal} {Rev. Mod. Phys.}\ }\textbf {\bibinfo {volume}
  {49}},\ \bibinfo {pages} {435} (\bibinfo {year} {1977})}\BibitemShut
  {NoStop}%
\bibitem [{\citenamefont {Jayannavar}(1991)}]{jayannavar1991brownian}%
  \BibitemOpen
  \bibfield  {author} {\bibinfo {author} {\bibfnamefont {A.}~\bibnamefont
  {Jayannavar}},\ }\href {\doibase 10.1007/BF01313998} {\bibfield  {journal}
  {\bibinfo  {journal} {Zeitschrift für Physik B Condensed Matter}\ }\textbf
  {\bibinfo {volume} {82}},\ \bibinfo {pages} {153} (\bibinfo {year}
  {1991})}\BibitemShut {NoStop}%
\bibitem [{\citenamefont {Garanin}\ \emph {et~al.}(1990)\citenamefont
  {Garanin}, \citenamefont {Ishchenko},\ and\ \citenamefont
  {Panina}}]{garanin1990dynamics}%
  \BibitemOpen
  \bibfield  {author} {\bibinfo {author} {\bibfnamefont {D.}~\bibnamefont
  {Garanin}}, \bibinfo {author} {\bibfnamefont {V.}~\bibnamefont {Ishchenko}},
  \ and\ \bibinfo {author} {\bibfnamefont {L.}~\bibnamefont {Panina}},\ }\href
  {\doibase 10.1007/BF01079045} {\bibfield  {journal} {\bibinfo  {journal}
  {Theoretical and Mathematical Physics}\ }\textbf {\bibinfo {volume} {82}},\
  \bibinfo {pages} {169} (\bibinfo {year} {1990})}\BibitemShut {NoStop}%
\bibitem [{\citenamefont {Garanin}(1991)}]{Garanin1991}%
  \BibitemOpen
  \bibfield  {author} {\bibinfo {author} {\bibfnamefont {D.}~\bibnamefont
  {Garanin}},\ }\href {\doibase http://dx.doi.org/10.1016/0378-4371(91)90395-S}
  {\bibfield  {journal} {\bibinfo  {journal} {Physica A: Statistical Mechanics
  and its Applications}\ }\textbf {\bibinfo {volume} {172}},\ \bibinfo {pages}
  {470 } (\bibinfo {year} {1991})}\BibitemShut {NoStop}%
\bibitem [{\citenamefont {Plefka}(1993)}]{plefka1993}%
  \BibitemOpen
  \bibfield  {author} {\bibinfo {author} {\bibfnamefont {T.}~\bibnamefont
  {Plefka}},\ }\href@noop {} {\bibfield  {journal} {\bibinfo  {journal}
  {Zeitschrift f{\"u}r Physik B Condensed Matter}\ }\textbf {\bibinfo {volume}
  {90}},\ \bibinfo {pages} {447} (\bibinfo {year} {1993})}\BibitemShut
  {NoStop}%
\bibitem [{\citenamefont {Garanin}(1997)}]{garanin1997fokker}%
  \BibitemOpen
  \bibfield  {author} {\bibinfo {author} {\bibfnamefont {D.~A.}\ \bibnamefont
  {Garanin}},\ }\href {\doibase 10.1103/PhysRevB.55.3050} {\bibfield  {journal}
  {\bibinfo  {journal} {Phys. Rev. B}\ }\textbf {\bibinfo {volume} {55}},\
  \bibinfo {pages} {3050} (\bibinfo {year} {1997})}\BibitemShut {NoStop}%
\bibitem [{\citenamefont {Nussinov}\ \emph {et~al.}(2005)\citenamefont
  {Nussinov}, \citenamefont {Shnirman}, \citenamefont {Arovas}, \citenamefont
  {Balatsky},\ and\ \citenamefont {Zhu}}]{nussinov2005spin}%
  \BibitemOpen
  \bibfield  {author} {\bibinfo {author} {\bibfnamefont {Z.}~\bibnamefont
  {Nussinov}}, \bibinfo {author} {\bibfnamefont {A.}~\bibnamefont {Shnirman}},
  \bibinfo {author} {\bibfnamefont {D.~P.}\ \bibnamefont {Arovas}}, \bibinfo
  {author} {\bibfnamefont {A.~V.}\ \bibnamefont {Balatsky}}, \ and\ \bibinfo
  {author} {\bibfnamefont {J.~X.}\ \bibnamefont {Zhu}},\ }\href {\doibase
  10.1103/PhysRevB.71.214520} {\bibfield  {journal} {\bibinfo  {journal} {Phys.
  Rev. B}\ }\textbf {\bibinfo {volume} {71}},\ \bibinfo {pages} {214520}
  (\bibinfo {year} {2005})}\BibitemShut {NoStop}%
\bibitem [{\citenamefont {Katsura}\ \emph {et~al.}(2006)\citenamefont
  {Katsura}, \citenamefont {Balatsky}, \citenamefont {Nussinov},\ and\
  \citenamefont {Nagaosa}}]{katsura2006voltage}%
  \BibitemOpen
  \bibfield  {author} {\bibinfo {author} {\bibfnamefont {H.}~\bibnamefont
  {Katsura}}, \bibinfo {author} {\bibfnamefont {A.~V.}\ \bibnamefont
  {Balatsky}}, \bibinfo {author} {\bibfnamefont {Z.}~\bibnamefont {Nussinov}},
  \ and\ \bibinfo {author} {\bibfnamefont {N.}~\bibnamefont {Nagaosa}},\ }\href
  {\doibase 10.1103/PhysRevB.73.212501} {\bibfield  {journal} {\bibinfo
  {journal} {Phys. Rev. B}\ }\textbf {\bibinfo {volume} {73}},\ \bibinfo
  {pages} {212501} (\bibinfo {year} {2006})}\BibitemShut {NoStop}%
\bibitem [{\citenamefont {Chudnovsky}\ \emph {et~al.}(2012)\citenamefont
  {Chudnovsky}, \citenamefont {Garanin},\ and\ \citenamefont
  {O’Keeffe}}]{chudnovsky2012conservation}%
  \BibitemOpen
  \bibfield  {author} {\bibinfo {author} {\bibfnamefont {E.}~\bibnamefont
  {Chudnovsky}}, \bibinfo {author} {\bibfnamefont {D.}~\bibnamefont {Garanin}},
  \ and\ \bibinfo {author} {\bibfnamefont {M.}~\bibnamefont {O’Keeffe}},\
  }\href {\doibase 10.1007/s10948-012-1410-y} {\bibfield  {journal} {\bibinfo
  {journal} {Journal of Superconductivity and Novel Magnetism}\ }\textbf
  {\bibinfo {volume} {25}},\ \bibinfo {pages} {1007} (\bibinfo {year}
  {2012})}\BibitemShut {NoStop}%
\bibitem [{\citenamefont {Boixo}\ \emph {et~al.}(2014)\citenamefont {Boixo},
  \citenamefont {R{\o}nnow}, \citenamefont {Isakov}, \citenamefont {Wang},
  \citenamefont {Wecker}, \citenamefont {Lidar}, \citenamefont {Martinis},\
  and\ \citenamefont {Troyer}}]{boixo2014}%
  \BibitemOpen
  \bibfield  {author} {\bibinfo {author} {\bibfnamefont {S.}~\bibnamefont
  {Boixo}}, \bibinfo {author} {\bibfnamefont {T.~F.}\ \bibnamefont
  {R{\o}nnow}}, \bibinfo {author} {\bibfnamefont {S.~V.}\ \bibnamefont
  {Isakov}}, \bibinfo {author} {\bibfnamefont {Z.}~\bibnamefont {Wang}},
  \bibinfo {author} {\bibfnamefont {D.}~\bibnamefont {Wecker}}, \bibinfo
  {author} {\bibfnamefont {D.~A.}\ \bibnamefont {Lidar}}, \bibinfo {author}
  {\bibfnamefont {J.~M.}\ \bibnamefont {Martinis}}, \ and\ \bibinfo {author}
  {\bibfnamefont {M.}~\bibnamefont {Troyer}},\ }\href {\doibase
  10.1038/nphys2900} {\bibfield  {journal} {\bibinfo  {journal} {Nature
  Physics}\ }\textbf {\bibinfo {volume} {10}},\ \bibinfo {pages} {218}
  (\bibinfo {year} {2014})}\BibitemShut {NoStop}%
\bibitem [{\citenamefont {Shin}\ \emph
  {et~al.}(2014{\natexlab{b}})\citenamefont {Shin}, \citenamefont {Smith},
  \citenamefont {Smolin},\ and\ \citenamefont {Vazirani}}]{shinagain2014}%
  \BibitemOpen
  \bibfield  {author} {\bibinfo {author} {\bibfnamefont {S.~W.}\ \bibnamefont
  {Shin}}, \bibinfo {author} {\bibfnamefont {G.}~\bibnamefont {Smith}},
  \bibinfo {author} {\bibfnamefont {J.~A.}\ \bibnamefont {Smolin}}, \ and\
  \bibinfo {author} {\bibfnamefont {U.}~\bibnamefont {Vazirani}},\ }\href
  {http://arxiv.org/abs/1404.6499} {\bibfield  {journal} {\bibinfo  {journal}
  {{arXiv}:1404.6499}\ } (\bibinfo {year} {2014}{\natexlab{b}})}\BibitemShut
  {NoStop}%
\bibitem [{\citenamefont {Crowley}\ \emph {et~al.}(2014)\citenamefont
  {Crowley}, \citenamefont {Duri{\ifmmode \acute{c}\else \'{c}\fi{}}},
  \citenamefont {Vinci}, \citenamefont {Warburton},\ and\ \citenamefont
  {Green}}]{crowley2014}%
  \BibitemOpen
  \bibfield  {author} {\bibinfo {author} {\bibfnamefont {P.~J.~D.}\
  \bibnamefont {Crowley}}, \bibinfo {author} {\bibfnamefont {T.}~\bibnamefont
  {Duri{\ifmmode \acute{c}\else \'{c}\fi{}}}}, \bibinfo {author} {\bibfnamefont
  {W.}~\bibnamefont {Vinci}}, \bibinfo {author} {\bibfnamefont {P.~A.}\
  \bibnamefont {Warburton}}, \ and\ \bibinfo {author} {\bibfnamefont {A.~G.}\
  \bibnamefont {Green}},\ }\href {\doibase 10.1103/PhysRevA.90.042317}
  {\bibfield  {journal} {\bibinfo  {journal} {Phys. Rev. A}\ }\textbf {\bibinfo
  {volume} {90}},\ \bibinfo {pages} {042317} (\bibinfo {year}
  {2014})}\BibitemShut {NoStop}%
\bibitem [{\citenamefont {Bauer}\ \emph {et~al.}(2015)\citenamefont {Bauer},
  \citenamefont {Wang}, \citenamefont {Pi{\v{z}}orn},\ and\ \citenamefont
  {Troyer}}]{bauer2015}%
  \BibitemOpen
  \bibfield  {author} {\bibinfo {author} {\bibfnamefont {B.}~\bibnamefont
  {Bauer}}, \bibinfo {author} {\bibfnamefont {L.}~\bibnamefont {Wang}},
  \bibinfo {author} {\bibfnamefont {I.}~\bibnamefont {Pi{\v{z}}orn}}, \ and\
  \bibinfo {author} {\bibfnamefont {M.}~\bibnamefont {Troyer}},\ }\href@noop {}
  {\bibfield  {journal} {\bibinfo  {journal} {arXiv preprint arXiv:1501.06914}\
  } (\bibinfo {year} {2015})}\BibitemShut {NoStop}%
\bibitem [{Note1()}]{Note1}%
  \BibitemOpen
  \bibinfo {note} {Neglecting zero-point fluctuations of the environment this
  reduces to the time domain form $\cexp {\eta _\alpha (t) \eta _\beta (t')} =
  2 \gamma T \delta _{\alpha \beta } \delta (t-t')$.}\BibitemShut {Stop}%
\bibitem [{\citenamefont {McDermott}(2009)}]{mcdermott2009}%
  \BibitemOpen
  \bibfield  {author} {\bibinfo {author} {\bibfnamefont {R.}~\bibnamefont
  {McDermott}},\ }\href {\doibase 10.1109/TASC.2008.2012255} {\bibfield
  {journal} {\bibinfo  {journal} {Applied Superconductivity, IEEE Transactions
  on}\ }\textbf {\bibinfo {volume} {19}},\ \bibinfo {pages} {2} (\bibinfo
  {year} {2009})}\BibitemShut {NoStop}%
\bibitem [{\citenamefont {Prokof'ev}\ and\ \citenamefont
  {Stamp}(2000)}]{prokof2000}%
  \BibitemOpen
  \bibfield  {author} {\bibinfo {author} {\bibfnamefont {N.}~\bibnamefont
  {Prokof'ev}}\ and\ \bibinfo {author} {\bibfnamefont {P.}~\bibnamefont
  {Stamp}},\ }\href@noop {} {\bibfield  {journal} {\bibinfo  {journal} {Reports
  on Progress in Physics}\ }\textbf {\bibinfo {volume} {63}},\ \bibinfo {pages}
  {669} (\bibinfo {year} {2000})}\BibitemShut {NoStop}%
\bibitem [{\citenamefont {Faoro}\ and\ \citenamefont
  {Ioffe}(2006)}]{faoro2006}%
  \BibitemOpen
  \bibfield  {author} {\bibinfo {author} {\bibfnamefont {L.}~\bibnamefont
  {Faoro}}\ and\ \bibinfo {author} {\bibfnamefont {L.~B.}\ \bibnamefont
  {Ioffe}},\ }\href {\doibase 10.1103/PhysRevLett.96.047001} {\bibfield
  {journal} {\bibinfo  {journal} {Phys. Rev. Lett.}\ }\textbf {\bibinfo
  {volume} {96}},\ \bibinfo {pages} {047001} (\bibinfo {year}
  {2006})}\BibitemShut {NoStop}%
\bibitem [{\citenamefont {de~Sousa}(2007)}]{desousa2007}%
  \BibitemOpen
  \bibfield  {author} {\bibinfo {author} {\bibfnamefont {R.}~\bibnamefont
  {de~Sousa}},\ }\href {\doibase 10.1103/PhysRevB.76.245306} {\bibfield
  {journal} {\bibinfo  {journal} {Phys. Rev. B}\ }\textbf {\bibinfo {volume}
  {76}},\ \bibinfo {pages} {245306} (\bibinfo {year} {2007})}\BibitemShut
  {NoStop}%
\bibitem [{\citenamefont {Shnirman}\ \emph {et~al.}(2002)\citenamefont
  {Shnirman}, \citenamefont {Makhlin},\ and\ \citenamefont
  {Sch{\"o}n}}]{shnirman2002}%
  \BibitemOpen
  \bibfield  {author} {\bibinfo {author} {\bibfnamefont {A.}~\bibnamefont
  {Shnirman}}, \bibinfo {author} {\bibfnamefont {Y.}~\bibnamefont {Makhlin}}, \
  and\ \bibinfo {author} {\bibfnamefont {G.}~\bibnamefont {Sch{\"o}n}},\
  }\href@noop {} {\bibfield  {journal} {\bibinfo  {journal} {Physica Scripta}\
  }\textbf {\bibinfo {volume} {2002}},\ \bibinfo {pages} {147} (\bibinfo {year}
  {2002})}\BibitemShut {NoStop}%
\bibitem [{\citenamefont {Dubé}\ and\ \citenamefont {Stamp}(2001)}]{dube2001}%
  \BibitemOpen
  \bibfield  {author} {\bibinfo {author} {\bibfnamefont {M.}~\bibnamefont
  {Dubé}}\ and\ \bibinfo {author} {\bibfnamefont {P.}~\bibnamefont {Stamp}},\
  }\href {\doibase http://dx.doi.org/10.1016/S0301-0104(01)00303-2} {\bibfield
  {journal} {\bibinfo  {journal} {Chemical Physics}\ }\textbf {\bibinfo
  {volume} {268}},\ \bibinfo {pages} {257 } (\bibinfo {year}
  {2001})}\BibitemShut {NoStop}%
\bibitem [{\citenamefont {Yoshihara}\ \emph {et~al.}(2006)\citenamefont
  {Yoshihara}, \citenamefont {Harrabi}, \citenamefont {Niskanen}, \citenamefont
  {Nakamura},\ and\ \citenamefont {Tsai}}]{yoshihara2006}%
  \BibitemOpen
  \bibfield  {author} {\bibinfo {author} {\bibfnamefont {F.}~\bibnamefont
  {Yoshihara}}, \bibinfo {author} {\bibfnamefont {K.}~\bibnamefont {Harrabi}},
  \bibinfo {author} {\bibfnamefont {A.~O.}\ \bibnamefont {Niskanen}}, \bibinfo
  {author} {\bibfnamefont {Y.}~\bibnamefont {Nakamura}}, \ and\ \bibinfo
  {author} {\bibfnamefont {J.~S.}\ \bibnamefont {Tsai}},\ }\href {\doibase
  10.1103/PhysRevLett.97.167001} {\bibfield  {journal} {\bibinfo  {journal}
  {Phys. Rev. Lett.}\ }\textbf {\bibinfo {volume} {97}},\ \bibinfo {pages}
  {167001} (\bibinfo {year} {2006})}\BibitemShut {NoStop}%
\bibitem [{\citenamefont {Amin}\ \emph {et~al.}(2009)\citenamefont {Amin},
  \citenamefont {Truncik},\ and\ \citenamefont {Averin}}]{Amin2009}%
  \BibitemOpen
  \bibfield  {author} {\bibinfo {author} {\bibfnamefont {M.~H.~S.}\
  \bibnamefont {Amin}}, \bibinfo {author} {\bibfnamefont {C.~J.~S.}\
  \bibnamefont {Truncik}}, \ and\ \bibinfo {author} {\bibfnamefont {D.~V.}\
  \bibnamefont {Averin}},\ }\href {\doibase 10.1103/PhysRevA.80.022303}
  {\bibfield  {journal} {\bibinfo  {journal} {Phys. Rev. A}\ }\textbf {\bibinfo
  {volume} {80}},\ \bibinfo {pages} {022303} (\bibinfo {year}
  {2009})}\BibitemShut {NoStop}%
\bibitem [{Note2()}]{Note2}%
  \BibitemOpen
  \bibinfo {note} {As shown in appendix~\ref {sec:FPE} these differences also
  exist at the ensemble level.}\BibitemShut {Stop}%
\bibitem [{\citenamefont {Garrahan}\ \emph {et~al.}(2009)\citenamefont
  {Garrahan}, \citenamefont {Jack}, \citenamefont {Lecomte}, \citenamefont
  {Pitard}, \citenamefont {van Duijvendijk},\ and\ \citenamefont {van
  Wijland}}]{garrahan2009first}%
  \BibitemOpen
  \bibfield  {author} {\bibinfo {author} {\bibfnamefont {J.~P.}\ \bibnamefont
  {Garrahan}}, \bibinfo {author} {\bibfnamefont {R.~L.}\ \bibnamefont {Jack}},
  \bibinfo {author} {\bibfnamefont {V.}~\bibnamefont {Lecomte}}, \bibinfo
  {author} {\bibfnamefont {E.}~\bibnamefont {Pitard}}, \bibinfo {author}
  {\bibfnamefont {K.}~\bibnamefont {van Duijvendijk}}, \ and\ \bibinfo {author}
  {\bibfnamefont {F.}~\bibnamefont {van Wijland}},\ }\href
  {http://stacks.iop.org/1751-8121/42/i=7/a=075007} {\bibfield  {journal}
  {\bibinfo  {journal} {Journal of Physics A: Mathematical and Theoretical}\
  }\textbf {\bibinfo {volume} {42}},\ \bibinfo {pages} {075007} (\bibinfo
  {year} {2009})}\BibitemShut {NoStop}%
\bibitem [{\citenamefont {Kikuchi}\ \emph {et~al.}(1991)\citenamefont
  {Kikuchi}, \citenamefont {Yoshida}, \citenamefont {Maekawa},\ and\
  \citenamefont {Watanabe}}]{kikuchi1991metropolis}%
  \BibitemOpen
  \bibfield  {author} {\bibinfo {author} {\bibfnamefont {K.}~\bibnamefont
  {Kikuchi}}, \bibinfo {author} {\bibfnamefont {M.}~\bibnamefont {Yoshida}},
  \bibinfo {author} {\bibfnamefont {T.}~\bibnamefont {Maekawa}}, \ and\
  \bibinfo {author} {\bibfnamefont {H.}~\bibnamefont {Watanabe}},\ }\href
  {\doibase http://dx.doi.org/10.1016/S0009-2614(91)85070-D} {\bibfield
  {journal} {\bibinfo  {journal} {Chemical Physics Letters}\ }\textbf {\bibinfo
  {volume} {185}},\ \bibinfo {pages} {335 } (\bibinfo {year}
  {1991})}\BibitemShut {NoStop}%
\bibitem [{\citenamefont {Nowak}\ \emph {et~al.}(2000)\citenamefont {Nowak},
  \citenamefont {Chantrell},\ and\ \citenamefont {Kennedy}}]{nowak2000monte}%
  \BibitemOpen
  \bibfield  {author} {\bibinfo {author} {\bibfnamefont {U.}~\bibnamefont
  {Nowak}}, \bibinfo {author} {\bibfnamefont {R.~W.}\ \bibnamefont
  {Chantrell}}, \ and\ \bibinfo {author} {\bibfnamefont {E.~C.}\ \bibnamefont
  {Kennedy}},\ }\href {\doibase 10.1103/PhysRevLett.84.163} {\bibfield
  {journal} {\bibinfo  {journal} {Phys. Rev. Lett.}\ }\textbf {\bibinfo
  {volume} {84}},\ \bibinfo {pages} {163} (\bibinfo {year} {2000})}\BibitemShut
  {NoStop}%
\bibitem [{\citenamefont {Cheng}\ \emph {et~al.}(2006)\citenamefont {Cheng},
  \citenamefont {Jalil}, \citenamefont {Lee},\ and\ \citenamefont
  {Okabe}}]{cheng2006mapping}%
  \BibitemOpen
  \bibfield  {author} {\bibinfo {author} {\bibfnamefont {X.~Z.}\ \bibnamefont
  {Cheng}}, \bibinfo {author} {\bibfnamefont {M.~B.~A.}\ \bibnamefont {Jalil}},
  \bibinfo {author} {\bibfnamefont {H.~K.}\ \bibnamefont {Lee}}, \ and\
  \bibinfo {author} {\bibfnamefont {Y.}~\bibnamefont {Okabe}},\ }\href
  {\doibase 10.1103/PhysRevLett.96.067208} {\bibfield  {journal} {\bibinfo
  {journal} {Phys. Rev. Lett.}\ }\textbf {\bibinfo {volume} {96}},\ \bibinfo
  {pages} {067208} (\bibinfo {year} {2006})}\BibitemShut {NoStop}%
\bibitem [{\citenamefont {Johnson}\ \emph {et~al.}(2011)\citenamefont
  {Johnson}, \citenamefont {Amin}, \citenamefont {Gildert}, \citenamefont
  {Lanting}, \citenamefont {Hamze}, \citenamefont {Dickson}, \citenamefont
  {Harris}, \citenamefont {Berkley}, \citenamefont {Johansson}, \citenamefont
  {Bunyk} \emph {et~al.}}]{johnson2011quantum}%
  \BibitemOpen
  \bibfield  {author} {\bibinfo {author} {\bibfnamefont {M.}~\bibnamefont
  {Johnson}}, \bibinfo {author} {\bibfnamefont {M.}~\bibnamefont {Amin}},
  \bibinfo {author} {\bibfnamefont {S.}~\bibnamefont {Gildert}}, \bibinfo
  {author} {\bibfnamefont {T.}~\bibnamefont {Lanting}}, \bibinfo {author}
  {\bibfnamefont {F.}~\bibnamefont {Hamze}}, \bibinfo {author} {\bibfnamefont
  {N.}~\bibnamefont {Dickson}}, \bibinfo {author} {\bibfnamefont
  {R.}~\bibnamefont {Harris}}, \bibinfo {author} {\bibfnamefont
  {A.}~\bibnamefont {Berkley}}, \bibinfo {author} {\bibfnamefont
  {J.}~\bibnamefont {Johansson}}, \bibinfo {author} {\bibfnamefont
  {P.}~\bibnamefont {Bunyk}},  \emph {et~al.},\ }\href@noop {} {\bibfield
  {journal} {\bibinfo  {journal} {Nature}\ }\textbf {\bibinfo {volume} {473}},\
  \bibinfo {pages} {194} (\bibinfo {year} {2011})}\BibitemShut {NoStop}%
\bibitem [{Note3()}]{Note3}%
  \BibitemOpen
  \bibinfo {note} {It can be derived in a spinwave-like expansion, integrating
  out the components that do not lie in the great circle traversed by the
  adiabatic minimum.}\BibitemShut {Stop}%
\bibitem [{\citenamefont {Boixo}\ \emph {et~al.}(2013)\citenamefont {Boixo},
  \citenamefont {Albash}, \citenamefont {Spedalieri}, \citenamefont
  {Chancellor},\ and\ \citenamefont {Lidar}}]{boixo2013}%
  \BibitemOpen
  \bibfield  {author} {\bibinfo {author} {\bibfnamefont {S.}~\bibnamefont
  {Boixo}}, \bibinfo {author} {\bibfnamefont {T.}~\bibnamefont {Albash}},
  \bibinfo {author} {\bibfnamefont {F.~M.}\ \bibnamefont {Spedalieri}},
  \bibinfo {author} {\bibfnamefont {N.}~\bibnamefont {Chancellor}}, \ and\
  \bibinfo {author} {\bibfnamefont {D.~A.}\ \bibnamefont {Lidar}},\ }\href@noop
  {} {\bibfield  {journal} {\bibinfo  {journal} {Nature communications}\
  }\textbf {\bibinfo {volume} {4}} (\bibinfo {year} {2013})}\BibitemShut
  {NoStop}%
\bibitem [{\citenamefont {Albash}\ \emph {et~al.}(2015)\citenamefont {Albash},
  \citenamefont {Rønnow}, \citenamefont {Troyer},\ and\ \citenamefont
  {Lidar}}]{albash2014}%
  \BibitemOpen
  \bibfield  {author} {\bibinfo {author} {\bibfnamefont {T.}~\bibnamefont
  {Albash}}, \bibinfo {author} {\bibfnamefont {T.}~\bibnamefont {Rønnow}},
  \bibinfo {author} {\bibfnamefont {M.}~\bibnamefont {Troyer}}, \ and\ \bibinfo
  {author} {\bibfnamefont {D.}~\bibnamefont {Lidar}},\ }\href {\doibase
  10.1140/epjst/e2015-02346-0} {\bibfield  {journal} {\bibinfo  {journal} {The
  European Physical Journal Special Topics}\ }\textbf {\bibinfo {volume}
  {224}},\ \bibinfo {pages} {111} (\bibinfo {year} {2015})}\BibitemShut
  {NoStop}%
\bibitem [{Note4()}]{Note4}%
  \BibitemOpen
  \bibinfo {note} {The relation of the Metropolis algorithm to Fokker-Planck
  diffusion has been noted previously in general terms in Ref~\protect
  \rev@citealpnum {kikuchi1991metropolis} whilst other Monte--Carlo methods
  have been specifically connected with Langevin type dynamics in Refs~\protect
  \rev@citealpnum {nowak2000monte} and~\protect \rev@citealpnum
  {cheng2006mapping}}\BibitemShut {NoStop}%
\bibitem [{\citenamefont {Lanting}\ \emph {et~al.}(2011)\citenamefont
  {Lanting}, \citenamefont {Amin}, \citenamefont {Johnson}, \citenamefont
  {Altomare}, \citenamefont {Berkley}, \citenamefont {Gildert}, \citenamefont
  {Harris}, \citenamefont {Johansson}, \citenamefont {Bunyk}, \citenamefont
  {Ladizinsky}, \citenamefont {Tolkacheva},\ and\ \citenamefont
  {Averin}}]{lanting2011probing}%
  \BibitemOpen
  \bibfield  {author} {\bibinfo {author} {\bibfnamefont {T.}~\bibnamefont
  {Lanting}}, \bibinfo {author} {\bibfnamefont {M.~H.~S.}\ \bibnamefont
  {Amin}}, \bibinfo {author} {\bibfnamefont {M.~W.}\ \bibnamefont {Johnson}},
  \bibinfo {author} {\bibfnamefont {F.}~\bibnamefont {Altomare}}, \bibinfo
  {author} {\bibfnamefont {A.~J.}\ \bibnamefont {Berkley}}, \bibinfo {author}
  {\bibfnamefont {S.}~\bibnamefont {Gildert}}, \bibinfo {author} {\bibfnamefont
  {R.}~\bibnamefont {Harris}}, \bibinfo {author} {\bibfnamefont
  {J.}~\bibnamefont {Johansson}}, \bibinfo {author} {\bibfnamefont
  {P.}~\bibnamefont {Bunyk}}, \bibinfo {author} {\bibfnamefont
  {E.}~\bibnamefont {Ladizinsky}}, \bibinfo {author} {\bibfnamefont
  {E.}~\bibnamefont {Tolkacheva}}, \ and\ \bibinfo {author} {\bibfnamefont
  {D.~V.}\ \bibnamefont {Averin}},\ }\href {\doibase
  10.1103/PhysRevB.83.180502} {\bibfield  {journal} {\bibinfo  {journal} {Phys.
  Rev. B}\ }\textbf {\bibinfo {volume} {83}},\ \bibinfo {pages} {180502}
  (\bibinfo {year} {2011})}\BibitemShut {NoStop}%
\bibitem [{\citenamefont {Harris}\ \emph {et~al.}(2010)\citenamefont {Harris},
  \citenamefont {Johansson}, \citenamefont {Berkley}, \citenamefont {Johnson},
  \citenamefont {Lanting}, \citenamefont {Han}, \citenamefont {Bunyk},
  \citenamefont {Ladizinsky}, \citenamefont {Oh}, \citenamefont {Perminov},
  \citenamefont {Tolkacheva}, \citenamefont {Uchaikin}, \citenamefont
  {Chapple}, \citenamefont {Enderud}, \citenamefont {Rich}, \citenamefont
  {Thom}, \citenamefont {Wang}, \citenamefont {Wilson},\ and\ \citenamefont
  {Rose}}]{harris2010experimental}%
  \BibitemOpen
  \bibfield  {author} {\bibinfo {author} {\bibfnamefont {R.}~\bibnamefont
  {Harris}}, \bibinfo {author} {\bibfnamefont {J.}~\bibnamefont {Johansson}},
  \bibinfo {author} {\bibfnamefont {A.~J.}\ \bibnamefont {Berkley}}, \bibinfo
  {author} {\bibfnamefont {M.~W.}\ \bibnamefont {Johnson}}, \bibinfo {author}
  {\bibfnamefont {T.}~\bibnamefont {Lanting}}, \bibinfo {author} {\bibfnamefont
  {S.}~\bibnamefont {Han}}, \bibinfo {author} {\bibfnamefont {P.}~\bibnamefont
  {Bunyk}}, \bibinfo {author} {\bibfnamefont {E.}~\bibnamefont {Ladizinsky}},
  \bibinfo {author} {\bibfnamefont {T.}~\bibnamefont {Oh}}, \bibinfo {author}
  {\bibfnamefont {I.}~\bibnamefont {Perminov}}, \bibinfo {author}
  {\bibfnamefont {E.}~\bibnamefont {Tolkacheva}}, \bibinfo {author}
  {\bibfnamefont {S.}~\bibnamefont {Uchaikin}}, \bibinfo {author}
  {\bibfnamefont {E.~M.}\ \bibnamefont {Chapple}}, \bibinfo {author}
  {\bibfnamefont {C.}~\bibnamefont {Enderud}}, \bibinfo {author} {\bibfnamefont
  {C.}~\bibnamefont {Rich}}, \bibinfo {author} {\bibfnamefont {M.}~\bibnamefont
  {Thom}}, \bibinfo {author} {\bibfnamefont {J.}~\bibnamefont {Wang}}, \bibinfo
  {author} {\bibfnamefont {B.}~\bibnamefont {Wilson}}, \ and\ \bibinfo {author}
  {\bibfnamefont {G.}~\bibnamefont {Rose}},\ }\href {\doibase
  10.1103/PhysRevB.81.134510} {\bibfield  {journal} {\bibinfo  {journal} {Phys.
  Rev. B}\ }\textbf {\bibinfo {volume} {81}},\ \bibinfo {pages} {134510}
  (\bibinfo {year} {2010})}\BibitemShut {NoStop}%
\end{thebibliography}%

\appendix
\section{Strongly dissipative dynamics}
\label{sec:SDD}
In this appendix we derive the effective $\mathrm{O}(2)$ dynamics given in
equations~\eqref{eq:LowTO2} and~\eqref{eq:HighTO2}, in which the Markovian
model features a confinement of $\phi$.
 In the dissipative limit
$\tau_p \gg \tau_d$ the spin dynamics are given by
eq~\eqref{eq:StochasticProcess}. To leading order in the long timescale, the
$\phi$ dynamics take the form
\begin{equation}
\begin{aligned}
\dot{\phi}&= - s \gamma B \sin \theta^* \sin \theta \sin \phi + \eta =
-\tau_\phi^{-1} \sin \phi + \eta.
\end{aligned}
\label{eq:StochasticProcessApprox}
\end{equation}
where as before $\tau_\phi^{-1} = \gamma s B \sin \theta \sin \theta^*
$. This corresponds to a Fokker--Planck equation
\begin{equation}
\label{eq:phiFPE}
\pdev{p}{t} = -\pdev{}{\phi} \left(\frac{\sin \phi}{\tau_\phi} - \gamma T
\pdev{}{\phi} \right) p.
\end{equation}
for an ensemble distribution $p$. The equilibrium solution to
eq~\eqref{eq:phiFPE} is given by
\begin{equation}
p = \frac{\exp \left( A \cos \phi\right)}{2 \pi I_0(A)},
\end{equation}
where $A=\left(\gamma T \tau_\phi\right)^{-1}$ and $I_n(\cdot)$ are
the modified Bessel functions of the first kind. Including the additional
sub-leading terms (those in eq~\eqref{eq:StochasticProcess}
that are missing from eq~\eqref{eq:StochasticProcessApprox}) leads to a
Fokker-Planck equation with no closed form solution. However, the
salient features are captured by making an appropriate shift to $\phi$ and
rescaling of $A$. This leads to an equilibrium solution
\begin{equation}
\label{eq:phidist}
\begin{aligned}
p &= \frac{\exp \left( A \cos (\phi-\phi^*)\right)}{2 \pi I_0(A)},\\
\tan \phi^* &= \frac{\sin(\theta-\theta^*)}{s
\gamma \sin^2 \theta \sin \theta^*	},
\end{aligned}
\end{equation}

where $A = \left(\gamma T \tau_\phi \cos \phi^* \right)^{-1}$, which deviates
from the exact solution only far from the distribution peak.
In terms of $\theta$, $A$ takes the form $A =  B \sqrt{s^2 \gamma^2
\sin^4 \theta \sin^2 \theta^* + \sin^2 \left(\theta - \theta^*\right)}/ \left(\gamma T
\sin \theta\right)$. 

Assuming a separation of timescales, the remaining $\theta$ dynamics can be
found by averaging eq~\eqref{eq:StochasticProcess} over the equilibrium
distribution of $\phi$ given in eq~\eqref{eq:phidist}. This leaves a single
equation giving the $\theta$ dynamics, 
\begin{equation}
\label{eq:thetadot}
\begin{aligned}
\dot{\theta} &= - B \sin \theta^* \, \cexp{\sin \phi} + \eta',
\end{aligned}
\end{equation}
where $\theta$ is subject to a drift term originating from the mean value of
the fast $\phi$ dynamics and a stochastic term which originates from the
fluctuations in the $\phi$ dynamics away from their mean value.
This approximation becomes accurate on timescales $t \gg \tau_\phi$.
Evaluating this exactly we obtain
\begin{equation}
\begin{aligned}
\cexp{\sin \phi} &= \int \mathrm{d} \phi \, p(\phi) \, \sin \phi = \sin \phi^*
\frac{I_1 (A)}{I_0 (A)} \\
\end{aligned}
\end{equation}
for the mean, whereas the noise term is defined by its mean $\cexp{\eta'(t)} =
0$ and covariance
\begin{widetext} 
\begin{equation}
\begin{aligned}
\cexp{\eta'(t)\eta'(t')} &= (B \sin \theta^*)^2 
\left(\cexp{\sin \phi(t) \sin \phi(t')} - \cexp{\sin \phi(t)}\cexp{\sin \phi(t')} \right)\\
& \approx (B \sin \theta^*)^2 \exp(-|t-t'|/\tau_\phi) \left(\cexp{\sin^2 \phi} - \cexp{\sin \phi}^2\right) \\
& \approx 2 \tau_\phi (B \sin \theta^*)^2 \delta\left(t-t'\right) \left(\cexp{\sin^2 \phi} - \cexp{\sin \phi}^2\right) \\
&= 2 \tau_\phi (B \sin \theta^*)^2 \delta\left(t-t'\right) \left[
\sin^2 \phi^* \left(1- \frac{I_1 (A)}{ A I_0 (A)} - \frac{I_1^2 (A)}{I_0^2
(A)}\right)+ \cos^2 \phi^* \frac{I_1(A)}{ A I_0 (A)} \right].\\
\end{aligned}
\end{equation}
\end{widetext}
We consider the limiting cases in which the above form simplifies:
\textbf{For $\bm{T \ll B}$}, we have $A \gg 1$ and $I_1(A)/I_0(A) = 1
- 1/(2 A), + O(A^{-2})$ which gives, to leading order
\begin{equation}
\label{eq:Lim1}
\begin{aligned}
\cexp{\sin \phi} &= \frac{\sin(\theta-\theta_0)}{s \gamma \sin^2 \theta \sin
\theta^*}, \\
\cexp{\eta'(t)\eta'(t')} &=  2 \gamma T (B \tau_\phi \sin \theta^*)^2 
 \delta(t-t').
\end{aligned}
\end{equation}
\textbf{For $\bm{T \gg B}$}, we have $A \ll 1$ and $I_1(A)/I_0(A) = A/2 +
O(A^2)$, which gives, to leading order in $\tau_\mathrm{d}/\tau_\mathrm{p}$,
\begin{equation}
\label{eq:Lim2}
\begin{aligned}
\cexp{\sin \phi} &= \frac{B \sin(\theta-\theta_0)}{2 T \gamma \sin \theta
},\\ 
\cexp{\eta'(t)\eta'(t')} &= \tau_\phi (B \sin \theta^*)^2 
 \delta(t-t')
\end{aligned}
\end{equation}
Substituting the values of eq~\eqref{eq:Lim1} and~\eqref{eq:Lim2} into
eq~\eqref{eq:thetadot} gives the forms eq~\eqref{eq:LowTO2}
and~\eqref{eq:HighTO2} respectively in the main body of the paper.
\section{Drude dynamics}
\label{sec:DD}
In this appendix we derive the dynamical equation of a single spin coupled to a
Drude bath and show that in an appropriate limit the long timescale $\theta$
dynamics are given by the Markovian equation~\eqref{eq:LowTO2} with $\phi$
remaining typically close to $\phi \approx 0$.
Using the Drude density of states and evaluating eqs~\eqref{eq:eta} and~\eqref{eq:gamma}, in the limit of $T \gg \omega_\mathrm{d}$, one obtains
$\cexp{\eta(t)\eta(t')} = T \gamma(t-t')= T \gamma \omega_\mathrm{d} \exp
\left(-\omega_\mathrm{d} |t-t'|\right)$. 
This particular noise $\eta(t)$ can be written in terms of a $\delta$-correlated
stochastic dummy variable $\eta'$ as
\begin{equation} 
\eta(t) = -\omega_\mathrm{d} \int_{-\infty}^t \d t'
\mathrm{e}^{-\omega_\mathrm{d}(t-t')} \eta'(t')
\end{equation}
which satisfies $\cexp{\eta'(t)\eta'(t')} = 2 \gamma T \delta(t-t')$. 
Rewriting eq~\eqref{eq:gen_dyn} the dynamics are obtained as
\begin{equation}
\begin{aligned}
\dot{\mathbf{s}}_i &= -\mathbf{s}_i \times \left(\mathbf{B}_i +
\mathbf{B}_\mathrm{diss}(t) \right)
\\
\mathbf{B}_\mathrm{diss}(t) &= - 
\omega_\mathrm{d} \int_{-\infty}^t \d t' \mathrm{e}^{-\omega_\mathrm{d}(t-t')}
\left(\gamma \dot{s}_{i,z}(t') - \eta' \right) \hat{\mathbf{z}}.
\end{aligned}
\end{equation}
Writing the second equation in its differential form we find
\begin{equation}
\begin{aligned}
\dot{\mathbf{B}}_\mathrm{diss} &= -\omega_\mathrm{d} \left(
\mathbf{B}_\mathrm{diss} + \gamma \dot{s}_{i,z}\hat{\mathbf{z}} - \eta'
 \hat{\mathbf{z}} \right)\\
 &= -\omega_\mathrm{d} \left( \mathbf{B}_\mathrm{diss} - \gamma \hat{\mathbf{z}}
 \left[ \hat{\mathbf{z}} \cdot \left( \mathbf{s}_i \times \mathbf{B}_i  \right)
 \right] - \eta'
 \hat{\mathbf{z}} \right)
 \end{aligned}
 \label{eq:appBdiss}
\end{equation}
where the second part follows by substituting $\dot{s}_{i,z}$ and noting that $
\hat{\mathbf{z}} \cdot \left(\mathbf{s}_i \times \mathbf{B}_\mathrm{diss}\right)
= 0$. These dynamics equate to equation~\eqref{eq:DLLG} after trivial
relabelling.

Separating eq~\eqref{eq:appBdiss} into the fast ($\phi,B_\mathrm{diss}^{(z)}$)
and slow ($\theta$) degrees of freedom and solving the linearised equations of
motion for $\phi$ and $B_\mathrm{diss}^{(z)}$ we find that 
\begin{equation}
\begin{split}
\dot{\theta} =& - B \sin \theta^* \, \cexp{\sin \phi} + \eta', \\
\cexp{\sin \phi}=&  \frac{\sin(\theta-\theta_0)}{s \gamma \sin^2 \theta \sin
\theta^*} \\ 
\cexp{\eta'(t)\eta'(t')} = & (B \sin \theta^*)^2 \mathrm{Cov}(t,t')
\end{split}
\end{equation}
as before in equation~\eqref{eq:Lim1}. However now there are long lasting
oscillations in the noise, characterised by $\mathrm{Cov}(t,t') =
\cexp{\sin \phi(t) \sin \phi(t')}-\cexp{\sin \phi(t)}\cexp{\sin \phi(t')}$.
Evaluating this one finds that
\begin{equation}
\label{eq:DrudeNoise}
\begin{split}
 \cexp{\eta'(t)\eta'(t')} = & \gamma T
\tau_\phi (B \sin \theta^*)^2 \mathrm{e}^{-\frac{\omega_\mathrm{d} |t-t'|}{2}}
\times \\ & \left[ \cos \left( \omega | t-t' | \right) + \frac{\omega_\mathrm{d}}{2 \omega} \sin \left( \omega | t-t' | \right) \right]
\end{split}
\end{equation}
where $\omega =
\sqrt{\frac{\omega_\mathrm{d}}{\tau_\phi}\left(1-\frac{\omega_\mathrm{d}\tau_\phi}{4}\right)}$ characterises the
oscillation frequency of the correlations. These fast oscillations can be seen in fig~\ref{fig:trajectory}c. 

As expected on timescales much longer that $1/\omega_\mathrm{d}$ the
covariance has the same value as with the previous cases
\begin{equation}
\begin{split}
\cexp{\eta'(t)\eta'(t')} \approx & 2 \gamma T (B \tau_\phi \sin \theta^*)^2 
 \delta(t-t'),\\
 = & \frac{2 T}{\gamma (s \sin \theta)^2} \delta(t-t').
 \end{split}
\end{equation}
thus on timescales much greater than the inverse 
bath frequency we recover the Markovian case (eqn~\eqref{eq:Lim1}) and the
$\theta$ dynamics of eqn~\eqref{eq:LowTO2} .

\section{Ensemble dynamics}
\label{sec:FPE}
In this appendix we study the dynamics of an ensemble of non-interacting to show
the appearance of the Markovian anisotropic dynamics at ensemble level. The ensemble dynamics are captured by a Fokker-Planck equation which is found to have similarly anisotropic dynamics, which can in turn be related to Metropolis Hastings dynamics.

We show that in the simplest case of  a non-interacting spins the ensemble dynamics do not agree with the Bloch
equations, this is clear as the anisotropic nature of the dynamics persists even after the
ensemble averaging. This result is perhaps surprising so we provide an explicit
derivation from the Fokker--Planck equation.

Within the first moment approximation the ensemble is described by the
probability distribution
\begin{equation}
\label{eq:prob}
p(\mathbf{s},t) = \frac1Z \mathrm{e}^{\bm{\xi}(t) \cdot \mathbf{s}}, \qquad Z =
\frac{4 \pi}{\xi} \sinh \left(\xi s\right).
\end{equation}
The first moment approximation in not appropriate when considering systems of
interacting spins as correlations between the trajectories of different spins
cannot be neglected. Thus for the purpose of deriving ensemble dynamics,
we consider a system of many spins, without interactions between them, each
acting under the influence of an external field $\mathbf{B}$ and coupled
anisotropically, along the $\hat{\mathbf{z}}$-direction only, to an ohmic bath.

For a specific realisation of the history of the bath, the dynamics are given by
the equation~\eqref{eq:zLLG2}. When we sum over the histories of the bath, the
dynamics are described by the evolution of the probability
distribution~\eqref{eq:prob}.

Following the approach of
Refs~\onlinecite{garanin1990dynamics,garanin1997fokker}\textemdash in which an
analogous calculation is performed for an isotropic bath coupling, i.e.
microscopic dynamics corresponding to equation~\eqref{eq:LLG2}\textemdash we find that the
evolution of $p(\mathbf{s},t)$is described by the Fokker-Planck equation
\begin{equation}
\label{eq:FPE}
\dot{p} = \nabla_\mathbf{s} \cdot \big[ \mathbf{s} \times \mathbf{B} +
\gamma \mathbf{s} \times \hat{\mathbf{z}} \left( \hat{\mathbf{z}} \cdot 
\left[ \mathbf{s} \times \left(\mathbf{B} - T \nabla_\mathbf{s} \right) \right]
\right) \big] p.
\end{equation}
By substituting equation~\eqref{eq:prob} into equation~\eqref{eq:FPE}
and integrating over $\mathbf{s}$, the dynamics of the parameter
$\bm{\xi}=\xi \hat{\bm{\xi}}$ are found to be governed by the equation
\begin{widetext}
\begin{equation}
\label{eq:xidyn}
\dot{\bm{\xi}} + \bm{\xi} \times \mathbf{B}+ \gamma s^2 \left(
\frac{s}{\sigma} - \frac{3}{\xi s}\right)
\bm{\xi}
\times
\hat{\mathbf{z}} \left[\hat{\mathbf{z}} \cdot \left( \bm{\xi} \times \mathbf{B}
\right) \right]+ \gamma T \left[ \left( \frac{\sigma}{\xi \sigma'} \hat{\bm{\xi}} \hat{\bm{\xi}}^T
\right) + \left( \bm{1} - \hat{\bm{\xi}} \hat{\bm{\xi}}^T
\right)\right]\left( \bm{1} - \hat{\mathbf{z}} \hat{\mathbf{z}}^T\right)\left(
\bm{\xi} - \bm{\xi}_0\right) = 0.
\end{equation}
\end{widetext}
The fixed
point of the dynamics is given by $\bm{\xi}_0 = \mathbf{B}/T$, which
corresponds to the Boltzmann distribution. $\bm{\sigma} =
\cexp{\mathbf{s}}$ is the mean polarisation vector with norm $\sigma =
\frac{1}{Z} \frac{\mathrm{d} Z}{\mathrm{d} \xi} = s \left(\coth \left( \xi s
\right) - \frac{1}{\xi s} \right)$ and $\sigma' = \frac{\mathrm{d} \sigma}{\mathrm{d} \xi}$ its derivative.

The physical origins of the first three terms are clear, as they
correspond to the respective terms of equation~\eqref{eq:zLLG2}. The last
term corresponds to longitudinal relaxation with a rate $\Gamma_1 = \gamma T$
and transverse relaxation with a rate $\Gamma_2 = \frac{\gamma T \sigma}{\xi
\sigma'}$. It should be noted, however, that the dynamics described by these
terms differs from the usual isotropic case due to the projecting out of the
component in the $\hat{\mathbf{z}}$ direction, this reflects the underlying
anisotropy of the coupling.

It is possible to re-write equation~\eqref{eq:xidyn} to give the dynamics
of the ensemble polarisation $\bm{\sigma}$, where $\xi$ is defined implicitly
by the form of $\sigma(\xi)$. Doing so, one obtains
\begin{widetext}
\begin{equation}
\dot{\bm{\sigma}} + \bm{\sigma} \times \mathbf{B}+ \gamma \frac{s^2}{\sigma^2} \left(
1 - \frac{3 \sigma}{\xi s^2}\right)
\bm{\sigma}
\times
\hat{\mathbf{z}} \left[\hat{\mathbf{z}} \cdot \left( \bm{\sigma} \times \mathbf{B}
\right) \right]+ \gamma T \left( \bm{1} - \hat{\mathbf{z}} \hat{\mathbf{z}}^T\right)\left(
\bm{\sigma} - \frac{\sigma}{\xi T} \mathbf{B}\right) = 0.
\end{equation}
\end{widetext}
For a highly polarised ensemble, $\sigma \approx s$, at low temperatures,
the ensemble dynamics converge upon the microscopic dynamics of
equation~\eqref{eq:zLLG2} showing that the unique behaviours described in the
body of the manuscript persist in the ensemble dynamics of the system.
\begin{figure}

\includegraphics[width=240pt,resolution=1000]{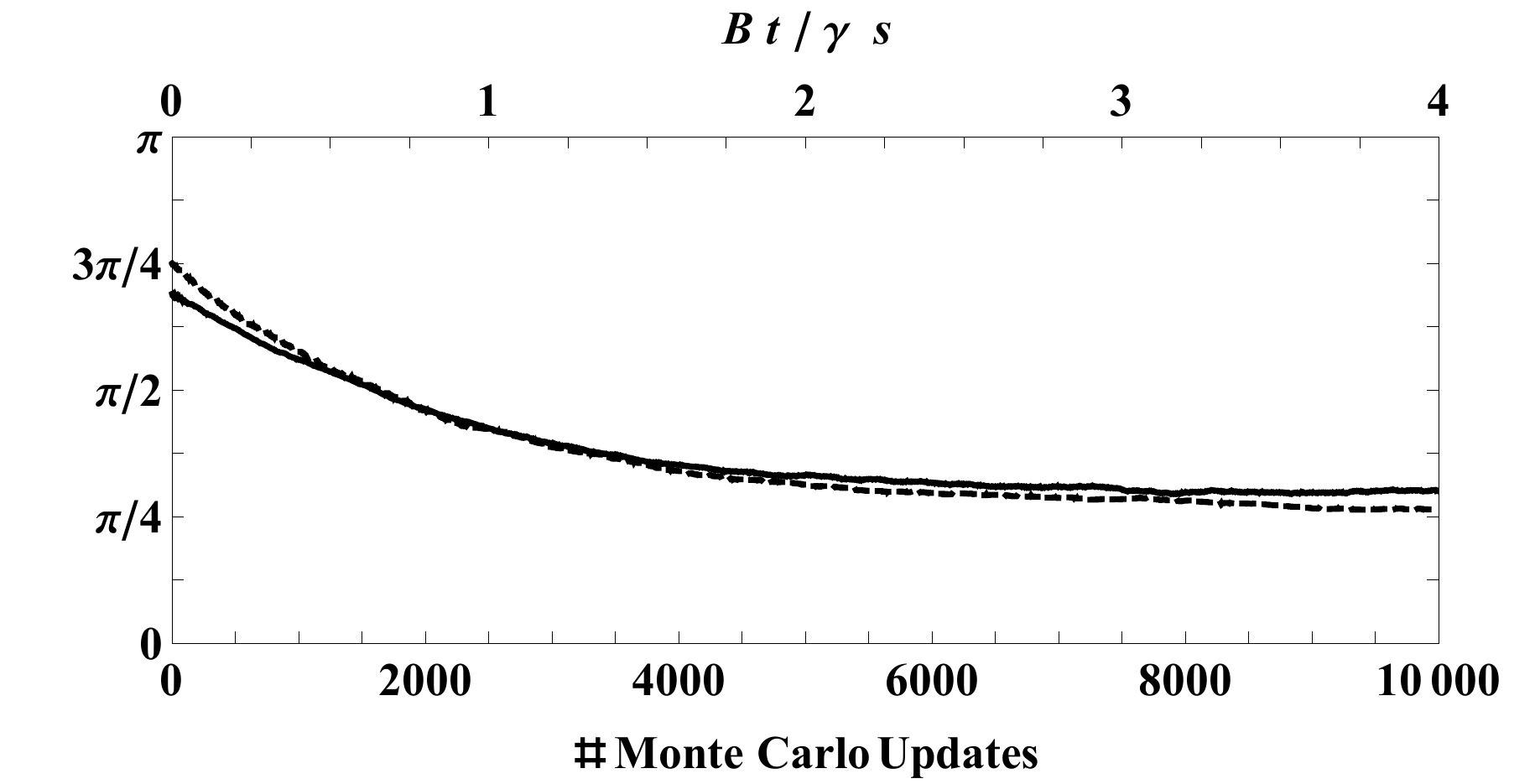}

\caption{
{\it Comparison of Metropolis Hastings and anisotropic LLG dynamics.} The ensemble averaged value $\cexp{\theta}$ of Metropolis Hastings (dashed) dynamics is compared with the anistropic LLG dynamics (solid) data from fig~\ref{fig:ensemble}. The LLG dynamics have small oscillations (see fig~\ref{fig:trajectory} for further detail), thus we plot the ensemble averaged mid-point of these small oscillations, $\cexp{\theta_M}$, as previously in fig~\ref{fig:ensemble}. Both simulation were performed using an ensemble of 1000 spins ($s=1/2$) initialised at $\theta = 3 \pi /4$, evolving in a magnetic field in the $\theta^* = \pi/4$, $\phi^* = 0$ direction, and a temperature $T = B/10$. For the LLG dynamics the coupling is $\gamma = 5\times10^3$, with energy scales $B = 10 T = 100 \omega_\mathrm{d}$, satisfying $B \gg T \gg \omega_\mathrm{d} \gg B / \gamma s$.
}
\label{fig:comparison}
\end{figure}
These ensemble dynamics are shown in fig~\ref{fig:ensemble} where they are obtained from sampling the stochastic sampling of the trajectories defined by eq~\ref{eq:zLLG0}, a sample of these trajectories is shown in fig~\ref{fig:trajectory}. In the main text we discuss the similarity between these dynamics and the Metropolis-Hastings dynamics of refs. ~\onlinecite{shin2014} and~\onlinecite{shinagain2014}. This similarity is most evident in their confinement to dynamics on an $\mathrm{O}(2)$ manifold, but is clear also in the similarity between their dynamics, both of which are overdamped and dissipative, as shown in fig~\ref{fig:comparison}. The existence of such a relationship between overdamped spin dynamics and the Metropolis Hastings algorithm has been previously established~\cite{kikuchi1991metropolis,nowak2000monte,cheng2006mapping}.

\end{document}